\documentclass[twocolumn,superscriptaddress,amsmath,amssymb,floatfix,nofootinbib,a4paper]{revtex4-1}

\usepackage[pdftex]{graphicx}
\usepackage{bm}
\usepackage{color}
\usepackage{times}
\usepackage{amsthm}
\usepackage{amsmath}
\usepackage{amssymb}
\usepackage{mathtools}
\usepackage{soul}
\usepackage{bbold}
\usepackage[squaren,Gray]{SIunits}
\usepackage{titlesec}

\newcommand\underrel[3][]{\mathrel{\mathop{#3}\limits_{%
      \ifx c#1\relax\mathclap{#2}\else#2\fi}}}

\def\rv{\mathbf{r}}

\titleformat*{\section}{\large\raggedright\bfseries}
\titleformat{\subsection}[runin]{\bfseries}{}{0em}{}[ --- ]
\titlespacing{\subsection}{0ex}{0ex}{1ex}

\def\rv{\mathbf{r}}
\def\nv{\mathbf{n}}
\def\pv{\mathbf{p}}

\newcommand{\bsf}[1]{\textsf{\textbf{#1}}}

\begin{document}

\title{Clustering and ordering in cell assemblies with generic  asymmetric aligning interactions}

\author{Thibault Bertrand}
\email[Electronic address: ]{t.bertrand@imperial.ac.uk}
\affiliation{Department of Mathematics, Imperial College London, South Kensington Campus, London SW7 2AZ, United Kingdom}
\author{Joseph d'Alessandro}
\affiliation{Institut Jacques Monod, CNRS UMR 7592, University Paris Diderot, Paris 75013, France}
\author{Ananyo Maitra}
\affiliation{Laboratoire Jean Perrin, UMR 8237 CNRS, Sorbonne Universit\'{e}, 75005 Paris, France}
\author{Shreyansh Jain}
\affiliation{Institut Jacques Monod, CNRS UMR 7592, University Paris Diderot, Paris 75013, France}
\author{Barbara Mercier}
\affiliation{Institut Jacques Monod, CNRS UMR 7592, University Paris Diderot, Paris 75013, France}
\author{Ren\'{e}-Marc M\`{e}ge}
\affiliation{Institut Jacques Monod, CNRS UMR 7592, University Paris Diderot, Paris 75013, France}
\author{Benoit Ladoux}
\email[Electronic address: ]{benoit.ladoux@ijm.fr}
\affiliation{Institut Jacques Monod, CNRS UMR 7592, University Paris Diderot, Paris 75013, France}
\author{Rapha\"{e}l Voituriez}
\email[Electronic address: ]{voiturie@lptmc.jussieu.fr}
\affiliation{Laboratoire Jean Perrin, UMR 8237 CNRS, Sorbonne Universit\'{e}, 75005 Paris, France}
\affiliation{Laboratoire de Physique Th\'{e}orique de la Mati\`{e}re Condens\'{e}e, UMR 7600 CNRS, Sorbonne Universit\'{e}, 75005 Paris, France}

\date{\today}


\begin{abstract}
Collective cell migration plays an essential role in various biological processes, such as development or cancer proliferation. While cell-cell interactions are clearly key determinants of collective cell migration -- in addition to  individual cells self-propulsion -- the physical mechanisms that control the emergence of cell clustering and collective cell migration are still poorly understood. In particular, observations have shown that binary cell-cell collisions generally lead to anti-alignement of cell polarities and separation of pairs -- a process called contact inhibition of locomotion (CIL), which is expected to disfavor the formation of large scale cell clusters with coherent motion. Here, we adopt a joint experimental and theoretical approach to determine the large scale dynamics of cell assemblies from elementary pairwise cell-cell interaction rules. We quantify experimentally binary cell-cell interactions and show that they can be captured by a minimal equilibrium-like pairwise asymmetric aligning interaction potential that reproduces the CIL phenomenology. We identify its symmetry class, build the corresponding active hydrodynamic theory and show on general grounds that such  asymmetric aligning interaction destroys large scale clustering and ordering, leading instead  to a liquid-like microphase of cell clusters of finite size and short lived polarity, or to a fully dispersed isotropic phase. Finally, this shows that CIL-like asymmetric interactions in cellular systems -- or general active systems -- control cluster sizes and polarity, and can prevent large scale coarsening and long range polarity, except in the singular regime of dense confluent systems.
\end{abstract}

\maketitle
\newpage

The emergence of collective, coordinated  migration  is a striking property of  eukaryotic cell collectives \cite{Friedl:2004aa,Friedl:2009aa,mayor-natrevmcb-2016,Hakim:2017aa}. It is observed in key biological  processes in vivo such as development \cite{McLennan:2012aa,Shellard:2018aa}, cancer proliferation \cite{Clark:2015aa,Palamidessi:2019aa} or wound healing \cite{cai-pnas-2016} and has now been reproduced in various in vitro set-ups \cite{Omelchenko:2003aa,Farooqui:2005aa,Poujade:2007uq,Trepat:2009aa,Vedula:2012aa,Ladoux:2017aa,Duclos:2018aa,jain-natphys-2020}.  This ability of cells to form large scale cohesive, polarized, self-propelled clusters is expected to be controlled both by single cells properties (polarity and motility), and  cell-cell interactions, as is confirmed experimentally \cite{Farhadifar:2007aa,Angelini:2010aa,Angelini:2011aa,Basan:2013aa,Sunyer:2016aa,Ladoux:2017aa}. 

\begin{figure*}[ht!]
\centering
\includegraphics[width=\textwidth]{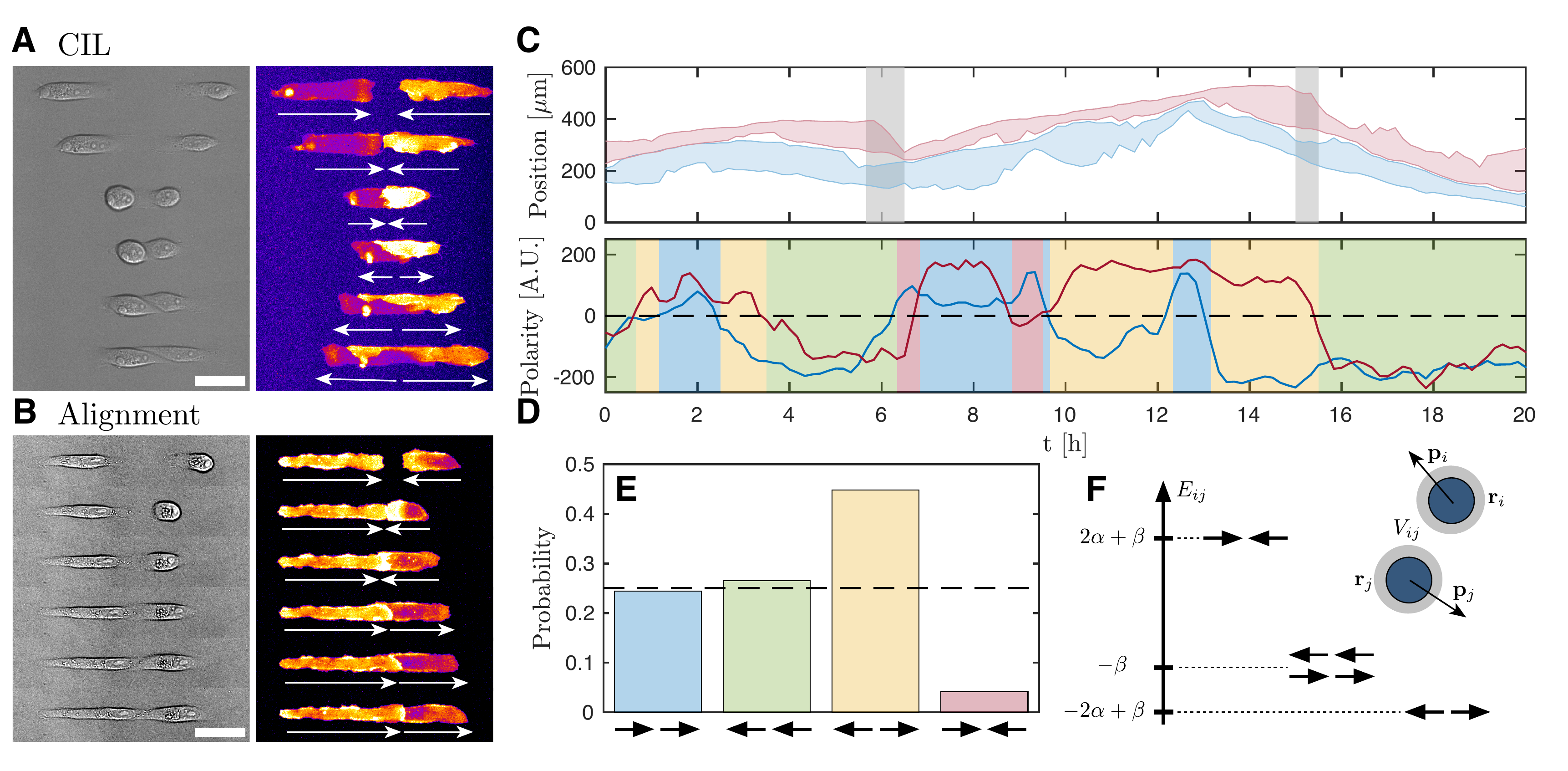}
\caption{{\bf Scattering rules in cell doublets} --- Potential outcomes of a collision between two cells (A) contact inhibition of locomotion (CIL) and (B) cell alignment shown in transmission (left) and in PBD-YFP fluorescence (right), in both panels time progresses from top to bottom with consecutive snapshots separated by $\Delta t = 10\,$min, scale bar $= 50~\micro$m. (C) Position of a cell doublet as a function of time; the shaded grey areas correspond to moments where the cells are not in contact.(D) Polarity of the cell doublet at corresponding times (see SI Appendix, Fig.S1). (E) Probabilities for the four possible polarity configurations; colors correspond to the highlighted regions in panel (C). (F) Schematic of the model with the energy levels $E_{ij}$ of the four possible doublet configurations with values of $\alpha/ \beta \approx 2$ obtained from the probabilities in panel (E).}
\label{fig:figure1}
\end{figure*}

The effects of cell-cell interactions on collective cell dynamics are two-fold. Upon contact, cells can engage transmembrane adhesion molecules (such as cadherins) to form junctions \cite{ladoux-nrmcb-2017}. On the one hand, these cell-cell junctions act as an effective attractive force  that opposes the separation of cell pairs and therefore favors cell clustering. On the other hand, it was found that cell-cell junctions also impact cell polarity; indeed, binary cell-cell interaction events are reported to typically favor outward pointing, anti-aligned polarities (see Fig.\,\ref{fig:figure1}) and ultimately separation of cell pairs. In the literature, this is generically known as contact inhibition of locomotion (CIL) \cite{Stramer:2017aa}, even though quantitative experimental analysis of the phenomenon remain sparse \cite{Desai:2013aa}. Thus far, the paradigm introduced by CIL, which favors anti-alignement and separation of pairs, therefore seems inconsistent with the observation of large scale cell clusters with coherent motion \cite{Desai:2013aa,smeets-pnas-2016,Zimmermann:2016aa}; reconciling these observations and more generally determining, from the knowledge of basic pairwise cell-cell interaction rules, the conditions of emergence of cell clustering and collective motion remains an outstanding open question.

Cell motility generically relies on the nonequilibrium dynamics of the actin/myosin system, driven by ATP hydrolysis; from a physics standpoint, this makes the cell a prototypical self-propelled particle (SPP)  and cell assemblies a striking example of active matter \cite{Toner:2005,marchetti-rmp-2013}. Active matter based models of collective cell migration, which involve -- explicit or implicit -- specific choices of cell-cell interaction rules  have flourished \cite{Camley:2017aa}; these can take various forms, from agent-based models \cite{Szabo:2006aa,Belmonte:2008aa,Henkes:2011aa,Basan:2013aa, Garcia:2015aa,smeets-pnas-2016,Zimmermann:2016aa}   and active vertex models \cite{Farhadifar:2007aa,Bi:2015aa,Bi:2016aa}  to active hydrodynamics models \cite{Saw:2017aa,Duclos:2018aa,Perez-Gonzalez:2019aa}  and phase fields models  \cite{Camley:2014aa}. They point, mostly through numerical simulations, to a broad variety of possible phases that can help interpret experimental observations. In particular, agent based models endowed with specific rules aiming at mimicking the CIL phenomenology have been proposed, and pointed to a rich phenomenology \cite{Desai:2013aa, smeets-pnas-2016,Zimmermann:2016aa}.

Even the simplest interactions between SPPs can, in fact, have striking consequences at the collective level. For instance, a simple pairwise aligning interaction between SPPs, as introduced by Vicsek {\it et al.} \cite{vicsek-prl-1995}, can lead to clustering and large scale collective motion in settings where long-range order and phase separation would be forbidden for systems at equilibrium \cite{Toner:2005,marchetti-rmp-2013}. Another purely nonequilibrium collective effect is the propensity of SPPs to cluster or undergo phase-separation in the presence of purely repulsive interactions  \cite{cates-arcmp-2015,tjhung-prx-2018}. A systematic exploration of the phase space of possible behaviors for more realistic models of cell assemblies with specific interaction rules is expected to lead to an ever-increasing complexity and therefore seems inaccessible. In this context, hydrodynamic theories, which are insensitive to specific microscopic choices but governed by symmetry properties and conservation rules \cite{Toner:2005,marchetti-rmp-2013} are promising candidates to provide unifying principles.

Here, we adopt a  joint experimental and theoretical approach that integrates quantitative multiscale {\it in vitro} data, numerical simulations and active hydrodynamic theory to determine the conditions of emergence of clustering and collective motion in cell assemblies. We make use of microfabricated one-dimensional {\it in vitro} environments to quantitatively characterize the onset of cell clustering and collective motion from the scale of cell pairs to the scale of large aggregates. We experimentally analyze pairwise cell-cell interactions and show that the observed CIL phenomenology  can be rationalized by a minimal equilibrium-like asymmetric aligning interaction potential whose symmetry class we identify.  Based on experimental observations, we combine such asymmetric aligning interaction with a classical short range attractive potential that mimic cell-cell junctions, build the corresponding active agent-based model and propose  a minimal  active hydrodynamic theory of this symmetry class. We show both experimentally and theoretically that the asymmetric aligning interaction can drastically lower the persistence of finite cell clusters and reduce their size. We demonstrate that in  the large system limit this can lead to a  transition between a dispersed (gas) isotropic phase and a liquid-like microphase of cell clusters of finite size and short lived polarity, which is critically controlled by both the strength of the asymmetric interaction and the cells' self-propulsion force.  Our results are applicable to general active systems of the same symmetry class, and show that CIL-like interactions can regulate cluster sizes and polarity, and in particular  prevent large scale coarsening and long range polarity, except in the singular regime of dense confluent systems. 

\section*{CIL as a pairwise equilibrium-like interaction potential in MDCK cells} We first aim to quantitatively characterize pairwise cell-cell interactions rules. At the scale of a pair of cells, contact inhibition of locomotion dictates distinctive intercellular dynamical rules that have been described qualitatively in the literature \cite{Stramer:2017aa}. Upon contact, cell-cell junctions are formed and trigger mechanotransduction signals leading to the repolarization of the two cells away from the contact location. To confirm this phenomenology, we study the statistics of the possible outcomes following binary cell collisions in a controlled quasi-1d cell migration geometry. We isolated MDCK (Madin-Darby canine kidney) cell doublets on fibronectin-coated linear strips of width $w = 20\micro$m, which we obtained by microcontact printing on polydimethylsiloxane  (PDMS) \cite{vedula-mcb-2014}.

Cells were treated with mitomycin C to prevent cell division and maintain cell numbers. To study the dynamics of front-rear cell polarization, we use a fluorescent biosensor (p21-activated kinase binding domain, PBD) of active {\it Rac1} and {\it Cdc42} \cite{machacek-nature-2009}. Prior to collision, incoming cells show clear signs of polarization, with asymmetric internal organization and shape, and a stable lamellipodium at the leading edge (see Fig.\,\ref{fig:figure1}A,B). Upon contact between the lamellipodia and formation of a doublet due to cell-cell adhesion, we typically observe inversions of the PBD gradient, signaling an inversion of cell polarity, with two possible outcomes: (i)  both cells repolarize away from the contact  (see Fig.\,\ref{fig:figure1}A) or (ii) only one of the two cells repolarizes away from the contact, leading to the alignment of  cell polarities  (see Fig.\,\ref{fig:figure1}B).

Using the sign of PBD gradient as a proxy for cell polarity over the lifetime of cell doublets, we extract both the position (and extension) of the two cells and their instantaneous polarity (see Fig.\,\ref{fig:figure1}C, SI Appendix, Fig.\,S1 and Movie S1). From these time series, we measure the respective probabilities of the four possible configurations of cell polarities resulting from an interaction event (see Fig.\,\ref{fig:figure1}D). We quantitatively confirm the contact inhibition hypothesis:  tail-tail configurations ($\leftarrow\rightarrow$) are strongly favored, while  head-head configurations ($\rightarrow\leftarrow$) are strongly disfavored. Symmetric configurations ($\leftarrow\leftarrow$ and $\rightarrow\rightarrow$) are equally probable as expected.

\section*{Minimal active particle model with asymmetric aligning interactions} Based on these quantitative observations, we build a  model of active Brownian particles (ABP) which includes minimal asymmetric interaction rules that describe the observed CIL phenomenology. The model is introduced in arbitrary space dimension $d$, but numerical simulations will be performed for $d=1$ to reproduce the experimental setup. Each cell is described as a particle with position $\rv_i$ and endowed with a polarization vector $\pv_i$ (a unit vector). We start from an equilibrium description and assume that cell-cell interactions result from a microscopic Hamiltonian given by
\begin{equation}
{\cal H} = \sum_{i,j} U_r(\rv_i,\rv_j) + U_p(\rv_i,\rv_j,\pv_i,\pv_j).
\label{eq:hamiltonian}
\end{equation}
The position dependent part that models cell-cell steric repulsion and adhesion can be taken to be the classical truncated Lennard-Jones potential 
\begin{equation}\label{Ur}
U_r = \begin{cases} 
4\varepsilon \left[ \left(\sigma/r_{ij}\right)^{12} - \left(\sigma/r_{ij}\right)^{6} \right] & r_{ij} \le r_c \\
0 & r_{ij} > r_c
\end{cases}
\end{equation}
where $\rv_{ij} = \rv_{i} - \rv_{j}$, $r_{ij} = |\rv_{ij}|$ and $r_c$ defines the range of interaction and $\sigma$ the particle size. Our main conclusions will be independent of this specific choice, which is used in numerical simulations for convenience. To build the polarization interaction potential $U_p$ we note that (for example, in $d=2$) the CIL interaction explicitly breaks the invariance under independent rotations of space and polarity vectors, which is preserved by classical aligning interactions $\propto \pv_i \cdot \pv_j$ characteristic of XY models or their active counterparts, the class of Vicsek-like models \cite{vicsek-prl-1995,Toner:2005,marchetti-rmp-2013}\footnote{Vicsek-like models break the invariance under independent rotations of space and polarity vectors, but only dynamically}. More explicitly, for fixed positions $\rv_{i},\rv_{j}$ of an interacting pair, the CIL phenomenology dictates that the system is not invariant under the symmetry $\pv_i, \pv_j\to -\pv_i, -\pv_j$, and therefore, cannot simply be described via a potential $\propto \pv_i \cdot \pv_j$.  Expanding in powers of $\rv_{ij}$ and $\pv_i, \pv_j$, the simplest term that breaks the invariance under independent rotations of space and polarity vectors, while being invariant under their \emph{joint} rotation, has the form $(\pv_i - \pv_j) \cdot  \rv_{ij}$. Without loss of generality, we therefore consider an interaction potential of the form
\begin{equation}\label{Up}
U_p = \begin{cases} 
- \beta \pv_i \cdot \pv_j   - \alpha (\pv_i - \pv_j) \cdot \nv_{ij} & r_{ij} \le r_c \\
0 & r_{ij} > r_c
\end{cases}
\end{equation}
where $\nv_{ij} = \rv_{ij} / r_{ij}$. This interaction potential is composed of two terms: (1) a {\it Vicsek or XY-like alignment term }with interaction strength $\beta$, which, by construction, is invariant under  independent rotations of space and polarity vectors and (2) an {\it asymmetric alignment term} of amplitude $\alpha$ which explicitly breaks this symmetry and reproduces the CIL phenomenology. Of note, potentials of  the same symmetry class have been considered in the study of liquid crystals \cite{dhakal-pre-2010}, but their effect in active matter systems have not been systematically examined.
 From the measure of the probabilities for each cell doublet configurations (see Fig.\,\ref{fig:figure1}E), we can estimate the value of the relative strength of the alignment interactions and find that our experiments lead to $\alpha / \beta  \approx 2$ leading to the polarization energy levels shown in Fig.\,\ref{fig:figure1}F (see SI Appendix).

\begin{figure*}[ht!]
\centering
\includegraphics[width=\textwidth]{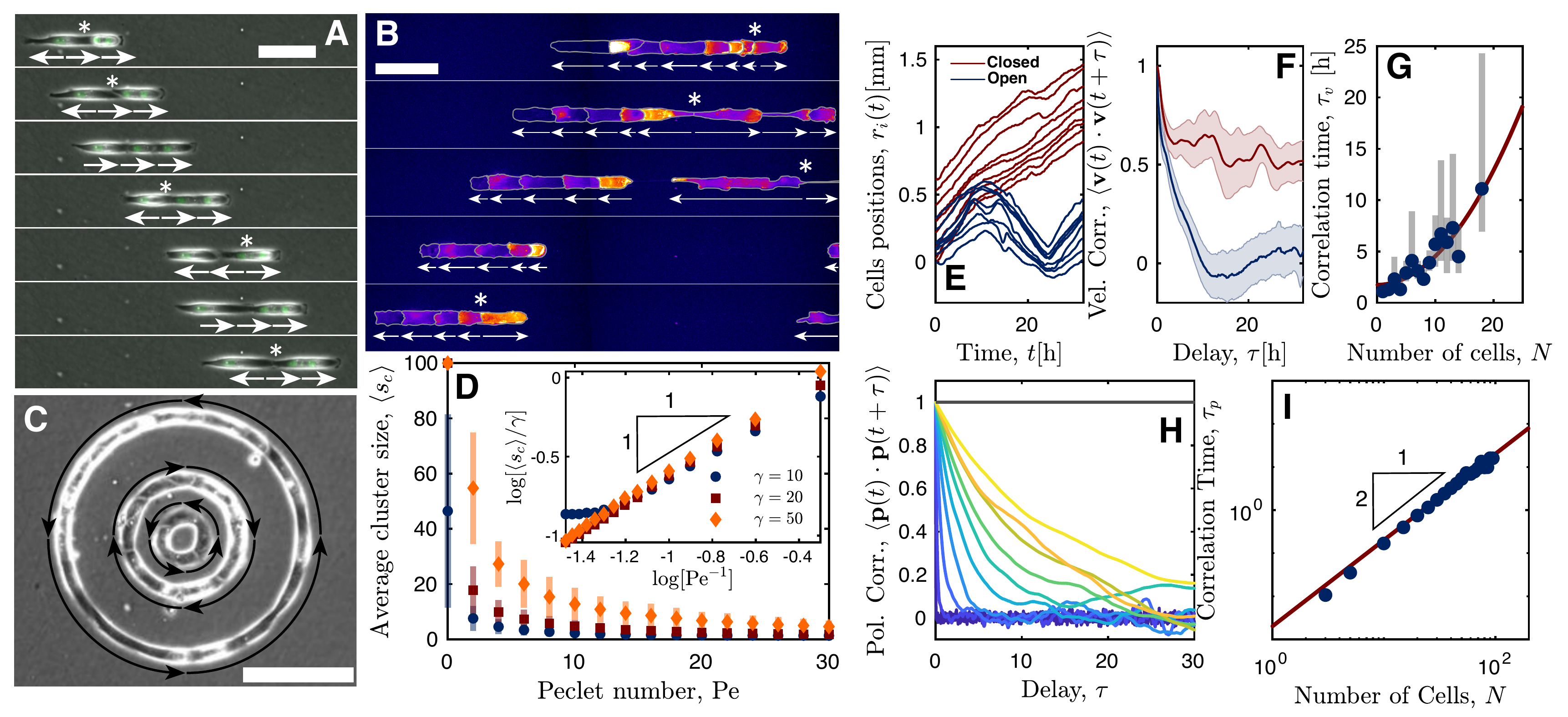}
\caption{{\bf Dynamics of small trains of cells} --- (A) Cohesive cell triplet showing anti-aligned polarities at the edges (arrows denote polarity and are provided as a guide) --  the center cell flips polarity in the middle slowing down the train. Snapshots are separated by $\Delta t = 120$ min, scale bar $= 100~\micro$m (see Movie S2). (B) Cell train containing 8 cells shown in  PBD-YFP fluorescence with repolarization of edge cells after fracture of the train; cell boundaries have been highlighted for clarity $\Delta t = 85$ min, scale bar $= 100 \micro$m (see Movie S3). In panels (A) and (B), the stars represent the location of the domain wall. (C) Examples of ring geometries at confluence (arrows show steady rotation direction for each ring); rings have diameters $D = 100\,\micro$m, $200\,\micro$m and $400\,\micro$m (scale bar $= 200\,\micro$m -- see Movie S4). (D) Steady state average cluster size $\langle s_c \rangle$ as a function of the P\'{e}clet number $\mathrm{Pe}$ for various cohesiveness $\gamma = 10$ (circles), $20$ (squares) and $50$ (diamonds) in the regime $\alpha \gg \beta$. Inset shows that $\langle s_c \rangle \propto (\mathrm{Pe}/\gamma)^{-1}$. (E) Position of cells in two trains for closed boundaries (ring geometry) and open boundaries (line geometry) showing high persistence in the case of closed boundaries. (F) Experimental velocity autocorrelation for open and closed boundaries. (G) Velocity correlation time (measured as the time for which correlation function reaches $1/e$) as a function of train size fitted by a quadratic law (red line). (H) Cell train global polarity autocorrelation function for various train sizes (increasing from blue to yellow -- $N \in [3,100]$) for open boundaries (for cohesive trains) and closed boundaries (horizontal grey line) averaged over 50 realizations. (I) Correlation time obtained from exponential fits of the autocorrelation function as a function of cell numbers $N$; the red solid line shows the theoretical prediction given by $\tau_p \sim N^2$. }
\label{fig:figure2}
\end{figure*}

Finally, this so-far equilibrium description is made minimally active by assuming that particles are subject to the self-propelling force $F_p$ along their polarization $\pv_i$. Our model can thus be interpreted as a generalization of the flying XY model \cite{Peruani:2008aa}, with a new asymmetric interaction term and a short-scale steric repulsion. More explicitly, for $d\ge2$, the dynamics of the system in the overdamped limit is governed by the set of coupled Langevin equations (see SI Appendix for $d=1$)
\begin{align}
\zeta \dot{\rv}_i & = - \frac{\partial {\cal H}}{\partial \rv_i} + F_p \pv_i + \sqrt{2T\zeta} \bm{\eta}_i  \label{eq:overdampedR}\\
\zeta_p \dot{\pv}_i & = ( \mathbb{1}- \pv_i \pv_i^{\intercal}) \cdot \left[ - \frac{\partial {\cal H} }{\partial \pv_i} + \sqrt{2T_p\zeta_p}\bm{\xi}_i \right] \label{eq:overdampedP}
\end{align}
where $\zeta$ is the friction coefficient, $\zeta_p$ is the rotational viscosity, $T,T_p$ are the translational and polarization temperatures (which can be different in out-of-equilibrium systems), $\bm{\eta}_i$ and $\bm{\xi}_i$ are zero mean and unit variance Gaussian white noises. The projection operator $\mathbb{1} - \pv_i \pv_i^{\intercal}$ ensures that the magnitude of $\pv_i$ remains invariant under the dynamics. It is useful to introduce the P\'{e}clet number ${\rm Pe}  = v_0 \sigma / D$, where $v_0 = F_p / \zeta$ is the self-propulsion velocity and $D = T / \zeta$ is the self-diffusion coefficient of the ABP. We will also make use of $\gamma = \varepsilon/T$ as the ratio of the strength of the Lennard-Jones potential to the thermal fluctuations, and similarly define $\alpha, \beta$ in units of $T_p$; finally, we introduce  the normalized relaxation rate $\mu=\tau/\zeta_p $ for the polarization, where $\tau = \sigma^2 / D$. Earlier agent based models \cite{Desai:2013aa, smeets-pnas-2016,Zimmermann:2016aa,george-scirep-2017} that take into account CIL interactions can be checked retrospectively to fall within this symmetry class, even though their specific choice of dynamics cannot be re-expressed as deriving from a simple pairwise effective potential.

The model is thus primarily controlled by (i) the volume fraction of particles $\phi$, (ii) the competition between self-propulsion ($\mathrm{Pe}$) and cohesion ($\gamma$), (iii) the strength of the symmetric and antisymmetric alignment interaction terms $\alpha,\beta$, and (iv) the relaxation rate $\mu$. Given this relative complexity, an exhaustive exploration of the phase behavior of this model goes beyond the scope of this paper. Below, we primarily aim to discuss the effect of the new asymmetric coupling $\alpha$ on the collective particle dynamics, and we restrict our analysis to regimes that are most relevant to our experimental cellular system.

\section*{Dynamics of  small cell trains} We first focus on the effect of the asymmetric interaction (parametrized by $\alpha$) on finite-sized cell clusters (or "trains''), based on our one-dimensional setup. As is shown in Fig.\,\ref{fig:figure2}A,B (and Movies S2 and S3), we observe that cells at the edges of cell trains generically have opposite polarities, pointing away from the center of mass of the train; this is expected from the CIL phenomenology -- as reported in Fig.\,\ref{fig:figure1} -- which favors $\leftarrow\rightarrow$ configurations. This is evidenced by the extension of lamellipodia (see Movie S3) and the PBD gradients (see Fig.\,\ref{fig:figure2}B). In a given cell train, we generally observe a single domain wall where the polarity changes sign ($\leftarrow\rightarrow$). This behavior can be simply accounted for by the interaction potential $U_p$ introduced in Eq.\,\ref{Up}. For $\alpha>\beta$, as is observed experimentally (see Fig.\,\ref{fig:figure1}D,E), the potential $U_p$ for a $1d$ train of $N$ particles is minimized for all configurations with a single domain wall $\leftarrow\rightarrow$;  in particular, inducing such domain wall in a fully polarized configuration leads to an energy gain $\Delta E=-2(\alpha-\beta)$. There is no energy cost incurred when the domain wall moves one step to the left or to the right in the bulk of the cell train (corresponding to a  polarization flip of a single particle); this suggests that domain walls perform symmetric random walks and thus diffuse  within the train. This directly results from the fact that the asymmetric interaction term in $U_p$ reduces to a boundary term, as is evident from summing the interaction potential over all particles of a finite $1d$ train of $N$ particles: 
\begin{equation}\label{Uptrain}
\sum_{i,j} U_p = 
- \beta \sum_{i=1}^{N-1} p_i p_{i+1}   + \alpha (p_1 - p_N),  \end{equation}
where $p_i=\pm1$ in $1d$. This shows that edges of cell clusters induce domain walls because of the CIL interactions. We  argue below that these CIL-induced domain walls  have important consequences on the dynamics of cell trains. 

\begin{figure*}[ht!]
\centering
\includegraphics[width=\textwidth]{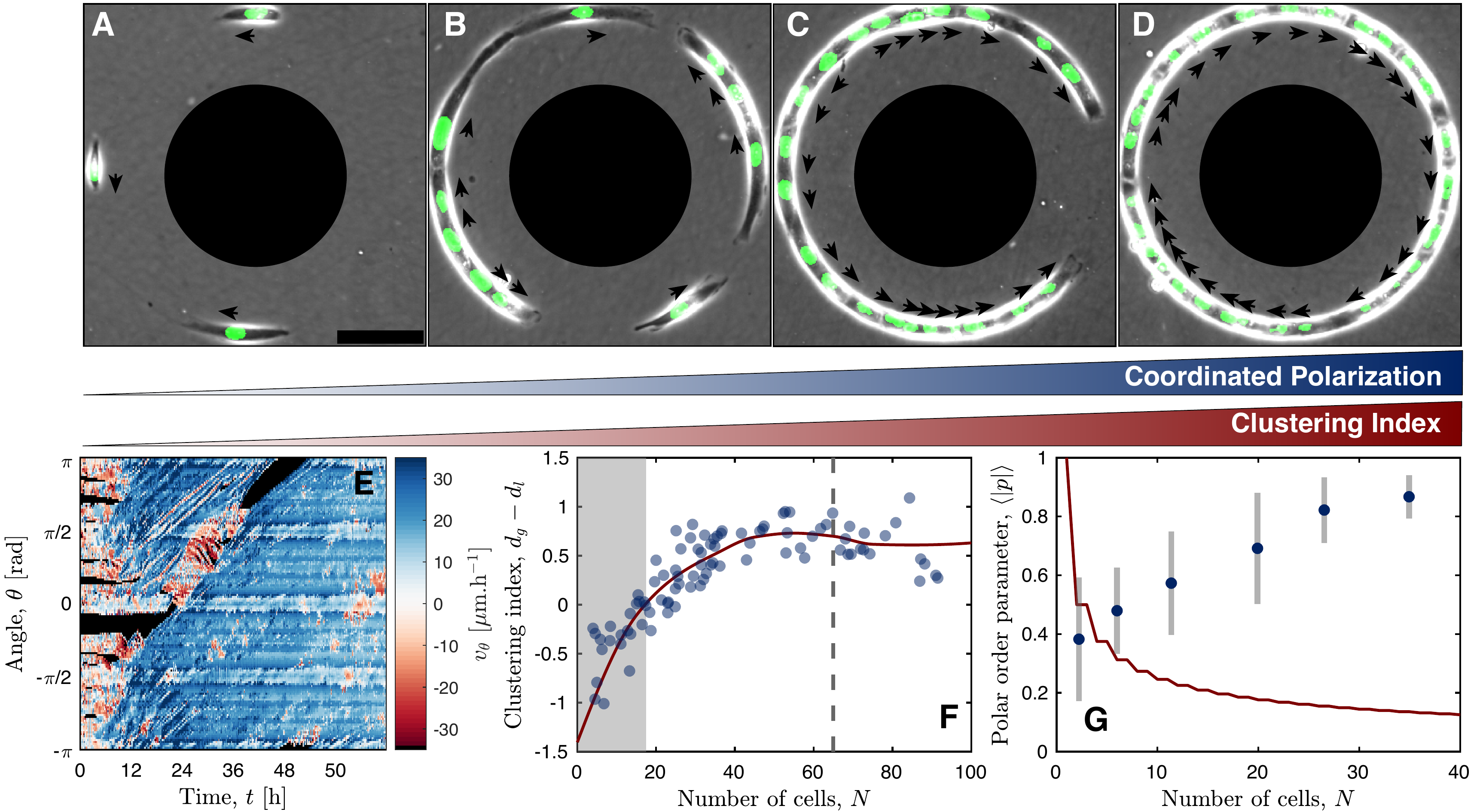}
\caption{{\bf Cell ordering and clustering in ring geometries} ---  (A-D) Example snapshots of systems with increasing number of cells in a ring geometry with fixed diameter $D = 400\,\micro$m; nuclei are fluorescently labelled, black arrows show the orientation of the velocity of the associated cell nucleus. As the number of cells increases (here, from $3$ to $28$ cells), the system reaches confluence; at confluence, cells coordinate their polarization and rotate in the same direction. [Scale bar $ = 100 \micro$m] (E) A kymograph of the orthoradial velocity $v_\theta$ (measured by PIV) as a function of time $t$ and angular position $\theta$ shows a transition to persistent rotational motion. The ring diameter is $D=1000\,\micro$m and cell number $N=91$. (F) Clustering index $d_g-d_l$ measured at long times $t > 48 \,\mathrm{h}$ as a function of the number of cells in the ring geometry with diameter $D = 1000\,\micro$m (blue circles). The data was binned and interpolated to produce the red line as a visual guide. The system transitions from a dispersed phase ($d_g-d_l<0$) to a clustered phase ($d_g-d_l>0$) for $N^*\approx20$, the vertical dashed line corresponds to the number of cells needed to reach confluence (see SI Appendix, Fig.S3). (G) Steady-state polar order parameter $\langle |p| \rangle$ averaged over realizations and time as a function of the number of cells in the ring geometry with diameter $D = 400\,\micro$m (error bars are given by standard deviations); the red line shows the expected value of the order parameter for a random set of $N$ polarizations (see SI Appendix).}
\label{fig:figure3}
\end{figure*}

First, Fig.\,\ref{fig:figure2}B shows that  domain walls can lead to the fragmentation of clusters, as expected from the outward pointing polarity and therefore propulsion force of cells on each side of the domain wall. Shortly after the separation, we observe that the cells at the newly formed edges repolarize away from the center of mass of their respective clusters, thereby inducing domain walls in the new clusters, confirming the above scenario of domain wall nucleation at cluster edges. This mechanism suggests that even for large cohesive interactions ($\gamma\gg1$, which would lead in equilibrium to clusters whose size diverges for $T\to0$ in $1d$), the asymmetric interaction can induce fragmentation of large clusters into smaller, finite-sized clusters in active systems. From force balance, we infer that only clusters of typical size $N\lesssim \gamma/ \mathrm{Pe}$ are insensitive to this active fragmentation mechanism. While extensive statistics of cluster sizes and controlled tuning of $\mathrm{Pe}$ and $\gamma$ are not accessible experimentally, we have verified that this scaling correctly predicts the average cluster size in numerical simulations of the $1d$ version of the model (see Fig.\,\ref{fig:figure2}D). Note that, as we discuss below, for increasing values of $\mathrm{Pe}$ a competing dynamic coarsening mechanism induced by the self propulsion of clusters occurs, and the proposed scaling is insufficient to capture the average cluster size. Finally, this shows quantitatively that the asymmetric interaction that we introduced can lead in active systems to a drastic reduction of cluster sizes, which is finite and critically controlled by $\mathrm{Pe}$.

Second, we now argue that domain walls in finite clusters control their dynamics, and, in particular, their self-propulsion speed and persistence. Defining the train polarity per cell as $p(t) = \frac{1}{N}\sum_{i}  p_i(t)$, we find from force balance that a train of $N$ cells is expected from the model to be self-propelled with (dimensionless) velocity $p(t)\mathrm{Pe}$. In turn, considering a train of $N$ cells with a single domain wall ($\alpha\gg\beta\gg 1 $ regime), the sign of $p(t)$ is determined by the relative position of the domain wall to the train center. Since domain walls diffuse in the model, we find that, starting from a random position in a train, a domain wall reaches the train center and thus induces a velocity sign change with a mean time $ \sim N^2$. We therefore expect that the polarity or velocity autocorrelation decays with a characteristic time $ \tau_p \sim N^2$ ; this is indeed clearly observed in both experimental data and numerical simulations (see Fig.\,\ref{fig:figure2}G,I).

Together, this shows that the CIL-based asymmetric interactions has striking consequences in active systems on size selection and dynamic properties of finite trains, as was suggested in   \cite{smeets-pnas-2016,Zimmermann:2016aa}. This, as we argued, is due to the nucleation of domain walls at the edges of cell trains. A very simple consequence of this analysis is that clusters with no edges, for example for dense confluent systems in closed periodic  geometries, should be completely insensitive to the CIL  asymmetric interaction. Strikingly, this is what we observed both experimentally and numerically when we analyzed periodic geometries (see Fig.\,\ref{fig:figure2}C, Fig.\,\ref{fig:figure3}A-D and Movie S4): for comparable system sizes, the persistence time was found in periodic geometries with cells at confluence to be significantly larger (larger than the observation time) than for trains with edges in open geometries (see Fig.\,\ref{fig:figure2}E,F). Indeed we observed that when the system reaches confluence, cluster edges disappear, domain walls vanish and sustained collective motion arises as predicted (see Fig.\,\ref{fig:figure3}A-D). Persistent collective rotational motion of confluent clusters in ring geometries was reported in \cite{jain-natphys-2020}, where the impact of CIL was however not discussed.

\section*{Impact of asymmetric interactions on collective cell behavior } Building on the previous analysis of finite cell trains, we now determine the impact of asymmetric interactions on the collective dynamics of large cell assemblies.  Experimentally, the only control parameter that is adjustable quantitatively is the cell density, which we tuned in ring geometries by varying the number of cells (see Fig.\,\ref{fig:figure3}A-D). Qualitatively, we observe  that an increase in cell density is associated with an increase of both cell-clustering and  coordinated polarization, as expected. As density increases, cells form larger clusters that are more persistent, in agreement with our analysis of single trains above. In Fig.\,\ref{fig:figure3}E, we clearly show the establishment of long-range coordinated motion in a large ring geometry at high density of cells. 

To make this analysis quantitative, we chose not to use the distribution of cluster sizes, because of the limited statistics that were accessible experimentally for each value $N$ of the cell number. Instead, in order to quantify the degree of clustering in the cell assembly, we defined two phenomenological parameters: $d_l$ measuring the effective distance between an experimental configuration and a perfect phase separated state  (i.e. a single cell aggregate) and $d_g$ measuring the distance between an experimental configuration and an ideal Poisson distribution  (i.e. a uniform distribution of non-overlapping cells) (see details in SI Appendix and Fig.S2). From these quantities, we introduce a clustering index $d_g - d_l$ which, by construction, differentiates between fully dispersed configurations ($d_g - d_l < 0$) and fully  clustered ones ($d_g - d_l > 0$). In Fig.\,\ref{fig:figure3}F, we show that this clustering index increases with $N$ (for a ring of diameter $D = 1000\micro$m), and plateaus beyond a value $N_c \approx 65$ when the system reaches confluence.  In turn, the polarity per cell $p$ introduced above can be accessed experimentally by identifying for each cell $p_i\equiv v_i/|v_i|$, where  $v_i$ is the velocity of cell $i$.  As discussed qualitatively above, we observe that the steady-state polarity averaged over time and realizations $\langle |p| \rangle$ increases with cell number $N$, which is consistent with the reported increase of cluster size with $N$, and the existence of typically one domain wall per cluster as reported in Fig.\,\ref{fig:figure2}; in particular, polarity significantly exceeds the expected value for a random choice of $N$ polarity vectors $p_i$, and plateaus for confluent systems with $N>N_c$ (see Fig.\,\ref{fig:figure3}G).

\begin{figure*}[ht!]
\centering
\includegraphics[width=\textwidth]{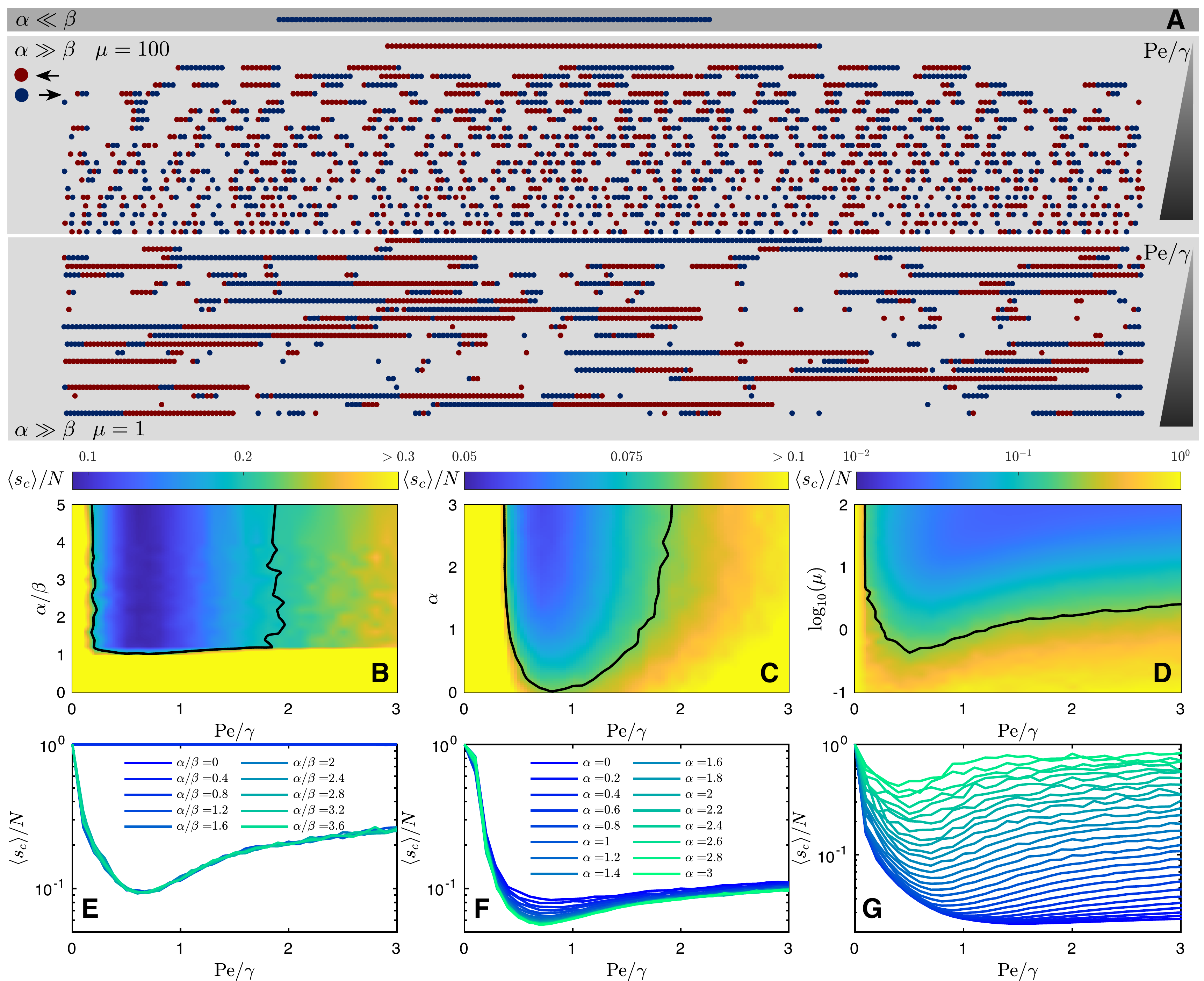}
\caption{ {\bf Polar-disordered and liquid-gas transitions in the model} ---  (A) Examples of steady-state structures observed in simulations of the model for various conditions $\alpha \ll \beta$ and $\alpha \gg \beta$ with $\mu = 100$ or $\mu = 1$. Cluster size statistics as a function of the control parameters of our model: the relative strength $\alpha/\beta$, the self-propulsion magnitude $\mathrm{Pe}/\gamma$ and the relaxation rate $\mu$ -- (B,E) in the ($\alpha/\beta$, ${\rm Pe}/\gamma$)-plane for $\mu = 100$; (C,F) in the ($\alpha$, ${\rm Pe}/\gamma$)-plane for $\mu = 2$ and $\beta \ll 1$; (D,G) in the ($\mu$, ${\rm Pe}/\gamma$)-plane for $\alpha/\beta \gg 1$. In panels (B-D), we have added as visual guides for the re-entrant clustering transition a single contour at arbitrary values of $\langle s_c \rangle /N$ (solid black line).}
\label{fig:figure4}
\end{figure*}

\section*{Hydrodynamic theory and phase diagram } Experiments are limited to rather small cell numbers, finite time scales and do not allow for the independent tuning of key control parameters. To explore the effect of asymmetric interactions on the possible phases in the thermodynamic limit, we rely on both the hydrodynamic limit of the agent-based model introduced above and on a numerical analysis of the $1d$ version of this model (see Fig.\,\ref{fig:figure4}A). 

We first construct a general hydrodynamic theory of asymmetrically interacting active polar particles in arbitrary dimensions, that encompasses the agent-based model introduced above (see details in SI Appendix). Since the particles interact with a substrate, momentum is not a conserved variable; hence, the hydrodynamic fields that we must retain are the particle density $\rho({\bf r},t)$, 
which is a conserved quantity and  the polarization ${\bf p}$. Note that even though the polarity ${\bf p}_i$ of a single particle is a unit vector, the corresponding coarse-grained variable ${\bf p}$ is not. The microscopic interaction potential ${\cal H}$ defined in Eq.\,\ref{eq:hamiltonian} can be coarse-grained to yield an effective free-energy in terms of the variables $\rho$ and ${\bf p}$. We do not perform this procedure explicitly here and instead provide a phenomenological expression based on symmetry arguments. The classical alignment  term  $-\beta{\bf p}_i\cdot{\bf p}_j$ in the microscopic polarity potential $U_p$ defined in Eq.\,\ref{Up} generically yields the usual Landau free energy \cite{gennes-book-1993}
\begin{equation}
F_p=\int d{\bf r}\left[\frac{A}{2}p^2+\frac{B}{4}p^4+\frac{K}{2}(\nabla{\bf p})^2\right]
\end{equation}
where all terms, in particular the parameter $A$ that controls the isotropic-polar transition, can be functions of the density. Note that the $(\nabla {\bf p})^2$ stands for the classical Frank elasticity in the one constant approximation, which comprises splay and bend contributions \cite{gennes-book-1993}.

As stated above, the asymmetric alignment term  $-\alpha({\bf p}_i-{\bf p}_j)\cdot{\bf n}_{ij}$ in Eq.\,\ref{Up}   breaks the invariance under independent rotations of space and polarity,  but preserves the required invariance under joint rotations of space and polarity that characterizes equilibrium polar liquid crystals. Upon coarse-graining, any asymmetric alignment term with such symmetry generically yields a coupling between density $\rho$ and space derivatives of ${\bf p}$; to lowest order in these hydrodynamic fields, we therefore write without loss of generality
\begin{equation}\label{F_p}
F_{\rho p} =-\int d{\bf r} \bar{\alpha}\delta\rho\nabla\cdot{\bf p}
\end{equation} 
where the coefficient $\bar{\alpha}$ is proportional to $\alpha$ and $\delta\rho$ is the deviation of the density from a steady-state, spatially homogeneous density $\rho_0$ (a free-energy $\propto\rho_0\nabla\cdot{\bf p}$ with a spatially homogeneous density $\rho_0$ would yield only a boundary term and is omitted). This is the spontaneous splay term, well known in the context of equilibrium polar liquid crystals \cite{Kung:2006aa}; in particular, it is apparent from Eq.\,\ref{F_p} that for finite particle clusters of uniform density, $F_{\rho p}$ is merely a boundary term, as we argued on the basis of the agent-based model above. As we show below this term has important consequences in active systems that have remained unexplored so far. 

Finally, the coarse-graining of the position dependent interaction potential $U_r$ yields
\begin{equation}
F_\rho=\int d{\bf r}\left[ U(\rho)+\frac{\kappa}{2}(\nabla\rho)^2\right]
\end{equation}
as in standard equilibrium theories of fluids where $U(\rho)$ is the internal energy density and $\kappa$ is the interface energy constant. Since we are interested in model independent properties, we choose a simple phenomenological form for $U(\rho)$ that accommodates a liquid-gas transition near a steady-state density $\rho_0$ controlled by a phenomenological parameter $A_c$, i.e. 
\begin{equation}
U(\rho)=\frac{A_c}{2}(\delta\rho)^2+\frac{B_c}{4}(\delta\rho)^4.
\end{equation}

\begin{figure*}[ht!]
\centering
\includegraphics[width=\textwidth]{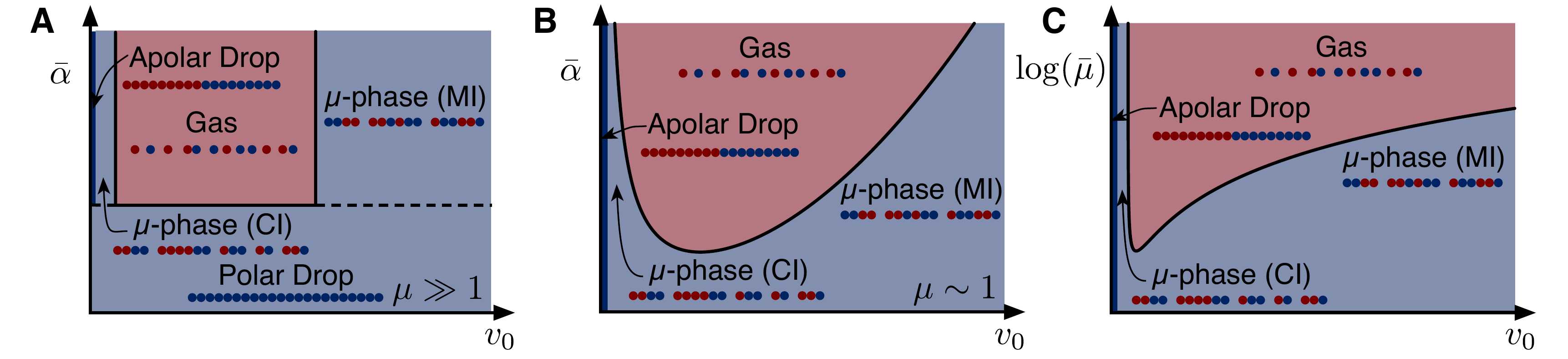}
\caption{ {\bf Phase diagram} --- represented respectively in (A) the ($\bar{\alpha},v_0$)-plane with $\bar{\mu} \gg 1$, (B) the ($\bar{\alpha},v_0$)-plane with $\bar{\mu} \sim 1$ and (C) the ($\bar{\mu},v_0$)-plane with $\bar{\alpha} \gg 1$. The black solid line in panels B-C is the spinodal line associated to Eq.(\ref{spinodal}) which defines the domain of linear stability of the homogeneous disordered phase (red region). Here, we assumed  that $\Lambda_2(v_0) $ is a decreasing quadratic function of $v_0$ with $\Lambda_2(v_0=0)=0$. All three phase diagrams clearly display re-rentrance to a clustered phase as the self-propelling velocity $v_0$ is increased.}
\label{fig:figure5}
\end{figure*}

The effect of the asymmetric contribution $F_{\rho p}$ to the total free energy $F=F_p+F_{\rho p}+F_\rho$ at equilibrium is summarized in SI Appendix for completeness, and shown to be unimportant for $d=1$, and  in the homogeneous \emph{disordered} phase for $d\ge 2$,  case that  we consider below. We now construct the phenomenological \emph{active} dynamics of $\rho$ and ${\bf p}$. The activity in the model enters primarily as a coarse-grained self-propulsion velocity of the polar particles that we assume enslaved to the polarity ${\bf v}\equiv v_0{\bf p}$. Building up on the classical model B for passive systems \cite{hohenberg-rmp-1977}, we write  the conservation equation for the particle density  as
\begin{equation}
\label{dens_pap}
\partial_t\rho=-\nabla\cdot(v_0\rho{\bf p})+D\nabla^2\frac{\delta F}{\delta \rho}+\Lambda_1\nabla\cdot\frac{\delta F}{\delta {\bf p}},
\end{equation}
while the polarization equation generically takes the form 
\begin{equation}
\label{pol_eq_pap}
\partial_t{\bf p}=-\bar{\mu}\frac{\delta F}{\delta {\bf p}}-\Lambda_2\nabla\frac{\delta F}{\delta \rho}.
\end{equation}
Here  $D$ is an effective diffusivity and $\bar{\mu}$ is the polarization relaxation rate which is the coarse-grained counter part of the parameter $\mu$ in the agent-based model. We have included phenomenological couplings between ${\bf p}$ and $\rho$ with coefficients $\Lambda_1$ and $\Lambda_2$. The equilibrium limit is defined by $v_0=0$ and $\Lambda_1=\Lambda_2$, imposed by Onsager symmetry, and corresponds to the linear response of a passive system of interacting polar particles. However, out of equilibrium, there is no symmetry to enforce this equality. Note that in principle, Eq.\,\ref{dens_pap} and Eq.\,\ref{pol_eq_pap} also contain nonlinear terms that cannot be derived from a potential (such as advective and self-advective nonlinear terms). These, however, do not affect the linear analysis below and are omitted for simplicity.

We now linearise Eq.\,\ref{dens_pap} and Eq.\,\ref{pol_eq_pap} about a disordered but homogeneous phase with density
$\rho_0$ (hence $A>0$). Upon elimination of the polarization field which relaxes fast, one obtains (see details in SI Appendix) :
\begin{equation}\label{spinodal}
\partial_t\rho=\left[v_0\rho_0\left(\frac{\bar{\alpha}}{A}+\frac{\Lambda_2A_c}{A\bar{\mu}}\right)-\frac{\Lambda_1\Lambda_2A_c}{\bar{\mu}}+A_cD\right]\nabla^2\rho.
\end{equation}
In equilibrium, i.e. for $v_0=0$ and $\Lambda_1=\Lambda_2$, the diffusivity is simply renormalized to $D-\Lambda_1^2/\bar{\mu}>0$ and the homogenous disordered state is linearly stable, as expected, for $A_c>0$; the usual spinodal line of the liquid-gas transition is then simply given by $A_c=0$.

At this stage, it is useful to compare this analysis with the equation of motion for the density and polarization fields in \cite{Geyer:2019aa}. These latter equations can be recovered in our formalism by taking $\Lambda_1=\bar{\alpha}=0$ and $\Lambda_2\not=0$. The condition $A_c(D+v_0\rho_0\Lambda_2/A\bar{\mu})<0$ is then equivalent to the motility induced phase separation (MIPS) \cite{cates-arcmp-2015,tjhung-prx-2018} spinodal obtained in \cite{Geyer:2019aa},  which arises in our notation when $\Lambda_2 A_c<0$, and which is indeed expected to be present in our system,  which has both self-propulsion and hard-core repulsion. To be consistent with \cite{Geyer:2019aa}, we therefore assume from now on that $\Lambda_1=0$, and that $A_c\Lambda_2(v_0) $ is a decreasing function of $v_0$ with $\Lambda_2(v_0=0)=0$. 

With these prescriptions, the spinodal line is explicitly determined by Eq.\,\ref{spinodal}, and defines the domain of linear stability of the homogeneous disordered phase, which we plot in the ($\bar{\alpha}$,$v_0$) and ($\bar{\mu}$,$v_0$) planes in Fig.\,\ref{fig:figure5}. This analysis however does not allow us to fully describe the expected phases beyond the spinodal lines. Therefore, we make use of numerical simulations of the $1d$ version of the agent-based model (see SI Appendix), which qualitatively confirm the analytical predictions and further characterize the inhomogeneous phases (Fig.\,\ref{fig:figure4}). Both analytical and numerical analysis show that the asymmetric term $\bar{\alpha}$, when coupled to activity ($v_0\not=0$) stabilizes the disordered homogeneous phase, both with respect to the usual, passive, mean-field transition to liquid induce by cohesive cellular forces  (the homogeneous gas phase can be stabilized even for $A_c<0$ when $\bar{\alpha}$ is increased) and with respect to MIPS (whose spinodal can be suppressed by increasing $\bar{\alpha}$). 

Strikingly, starting deep in the liquid phase ($A_c<0$) at equilibrium ($v_0=0$) induced by cohesive forces (CI), we find that for $\bar{\alpha}>0$, increasing self-propulsion first destabilizes clusters and induces a microphase of finite self propelled clusters, which can ultimately lead to a fully dispersed phase. This is in agreement with the mechanism of cluster fragmentation analysed in Fig.\,\ref{fig:figure2}B, and proves its impact at larger scales. For this mechanism to occur,  the asymmetric interactions have to dominate over the aligning ones ($\alpha>\beta$)
for $\beta>1$, or over the thermal fluctuations ($\alpha>1$) for $\beta<1$ (where $\alpha$ and $\beta$ are expressed in units of $T_p$). In addition to this fragmentation mechanism, we show in Fig.\,\ref{fig:figure3} that activity increases self-propulsion and persistence of clusters, which thus behave as mesoscale self-propelled particles with hard-core repulsion. Upon increasing activity, a competing MIPS mechanism of either single particles or clusters,  favoring aggregation is therefore expected. Such re-entrance into a clustered phase is indeed predicted by our stability analysis and numerical simulations (Fig.\,\ref{fig:figure4}). This MIPS induced (MI) clustering mechanism   is expected to be critically controlled by the persistence time of the clusters, which is in turn controlled by the time scale $\mu^{-1} $ of polarization dynamics, as confirmed by the linear stability analysis --- see the term $\rho_0 v_0 A_c \Lambda_2/(A\bar{\mu})$ in Eq.\,\ref{spinodal}. Indeed, we find that for large values of $\mu$, the MIPS phase is pushed to larger values of $v_0$, thus increasing the stability domain of the disordered dispersed phase; accordingly, for low values of $\mu$, the MIPS phase (MI) appears even at low $v_0$, thereby completely masking the fragmentation mechanism and preventing the emergence of a microphase (Figs.\,\ref{fig:figure4},\ref{fig:figure5}). 

\section*{Conclusion } Our results illustrate how cell-cell interactions regulate the collective behavior of cellular systems and their organization. Based on a joint experimental and theoretical approach, we analyzed the impact of generic asymmetric interactions reminiscent of the CIL interactions reported for various cell types on the collective dynamics of cell assemblies, and more generally of dry active systems. We made use of microfabricated one-dimensional {\it in vitro} environments to characterize quantitatively pairwise cell-cell interactions, and showed that the observed CIL-type phenomenology can be captured by a generic equilibrium-like asymmetric aligning interaction potential that breaks the usual invariance under {\it independent} rotations of space and polarities. Based on experimental observations, numerical simulations and analysis of the relevant active hydrodynamic theory, we demonstrated that such an asymmetric aligning interaction can drastically lower the size of cell clusters, and control their self-propulsion speed and persistence. In the large system limit, we found  that this can lead to the emergence of a liquid-like microphase of cell clusters of finite size and short lived polarity, and ultimately stabilize a fully dispersed  apolar phase. Altogether, this analysis  suggests that CIL-like asymmetric interactions in generic active systems -- cellular or artificial -- can control cluster sizes and polarity,  and can thus prevent large scale coarsening and long-ranged polarity, except in the singular regime of dense confluent systems. While our experimental and numerical analysis was focused on 1D geometries, we expect that several key features, predicted by the generic analysis of the $d$-dimensional hydrodynamic theory, are still valid for $d=2,3$, and therefore relevant to {\it in vivo} biological systems. We anticipate that the mechanism of active cluster fragmentation induced by the CIL interaction, and the critical dependence of size, speed and persistence of cell clusters on the CIL interaction may provide a new mechanism to interpret directed cell migration during development, epithelial-mesenchymal transition, and collective cancer cell invasion.

\begin{acknowledgements}
We acknowledge the ImagoSeine core facility of the Institut Jacques Monod, and members of IBiSA and France-BioImaging (ANR-10-INBS-04) infrastructures. The authors are grateful to F. Martin-Belmonte for their generous gift of MDCK cell lines. We also acknowledge financial support from the Agence Nationale de la Recherche (ANR) ``POLCAM" (ANR-17- CE13-0013) (RMM, BL, RV), the LABEX ``Who am I?" (ANR-11-LABX-0071), Universit\'{e} de Paris IdEx ANR-18-IDEX-0001, the Ligue Contre le Cancer (Equipe labellis\'{e}e) (BL, RMM), D\'{e}fi M\'{e}canobio CNRS MECAPOL 2016 (RMM, RV) and Agence Nationale de la Recherche (ANR) 'MECANADIPO' (ANR-17- CE13-XXXX) (BL). RV also acknowledges support from Fondation pour la Recherche M\'{e}dicale (FRM) and Institut National du Cancer (INCA).
\end{acknowledgements}

\section*{Author Contributions}

TB and JdA contributed equally to this work. TB, JdA, BL, and RV designed the research. BL and RV supervised the project. TB, AM and RV developed the theoretical model and numerical simulations. JdA, BM and SJ performed the experiments. TB, JdA, RMM, BL and RV analyzed the data.


%

\onecolumngrid
\pagebreak
\begin{center}
\textbf{\large Supplementary Material for "Clustering and ordering in cell assemblies with generic  asymmetric aligning interactions"}
\end{center}
\setcounter{equation}{0}
\setcounter{figure}{0}
\setcounter{table}{0}
\setcounter{page}{1}
\makeatletter
\renewcommand{\theequation}{S\arabic{equation}}
\renewcommand{\thefigure}{S\arabic{figure}}

\section*{Experimental methods}

\subsection*{Cell culture}
We used MDCK wild-type, MDCK histon-GFP and MDCK PBD-YFP (gift from F. Martin-Belmonte lab). The cells were cultured in DMEM  GlutaMAX high-glucose (Gibco, Waltham, MA) supplemented with 10\% f\oe tal bovine serum (BioWest, Nuaill\'e, France). Prior to experiments, the cells were treated with mitomycin C at final concentration of $10\,\micro\gram.\milli\liter$ added in the medium for $1\,\hour$, then rinsed before subsequent detachment and seeding on the experimental samples.
\\

\subsection*{Sample preparation} All micropatterns were prepared using standard micro-contact printing on PDMS, as described in Ref.\,\cite{peyret-biophysj-2019}. The substrates used were: (i) non-culture treated plastic dishes (Greiner Bio-One, Kremsm\"unster, Austria) for wide-field microscopy or (ii) glass coverslips (Menzel-Gl\"aser) for spinning-disk microscopy. The substrates were first covered with a thin layer of poly-dimethyl-syloxane (PDMS, Sylgard, Dow Corning, Midland, MI) using a spin-coater and crosslinked at $80\celsius$ for $2\,\hour$. PDMS stamps were made by pouring PDMS on a mold featuring the patterns to be printed and crosslinked as described. After cooling down, a fibronectin solution was prepared by adding $5\,\milli\gram.\milli\liter^{-1}$ of fibronectin and $2.5\,\milli\gram.\milli\liter^{-1}$ of Cy3- or Cy5-labelled fibronectin into sterile milliQ water. The solution was then incubated on the stamps for $40\,\minute$ at room temperature. Before stamping, the substrates were activated using UV-ozone for $10\,\minute$; the stamps were rinsed to remove any excess fibronectin and dried using an air-gun. The stamps were briefly put in contact with the surface of the substrate, then removed, and the substrates immerged in a $2\%$ pluronics F127 (Sigma-Aldrich, Saint Louis, MO) solution in PBS for $2\,\hour$. Finally, the substrates were rinsed in PBS and sterilized under the UV lamp of a culture hood before use.
\\

\subsection*{Stencils} To be able to track cell trains of definite length over a long period of time, we prepared isolated trains as follows. We fabricated PDMS micro-stencils by cutting a trapezoidal shape through a thin (approx. $100\,\micro\meter$) layer of PDMS using a cutting plotter (Graphtec CE6000-40, Graphtec Corp., Yokohama, Japan). The stencil was then placed on top of the linear patterns at a $90^\circ$ angle in order to leave from a few dozens of microns up to $400\,\micro\meter$ of uncovered space at the middle of each line. The stencil was removed after the standard cell seeding, attachment and rinsing steps, so as to let the trains move freely without future encounters.
\\
\subsection*{Cell seeding} The cells were enzymatically detached, then concentrated using a centrifuge and seeded on the substrates in culture medium, at a controlled density: medium-low density for the doublet and random small trains experiments, very high density for the stencil experiments, a wide range of densities for the ring experiments. The cells were let to adhere in the incubator for approx. $45\,\minute$, then rinsed thoroughly (but carefully) to remove excess floating cells without affecting adhered cells.
\\
\subsection*{Time-lapse microscopy} All experiments were run at 37\celsius\,in $5\%$ CO2. The experiments in Fig.\,1 and Fig.\,2B were done using an inverted microscope (Leica, Wetzlar, Germany) with a CSU-W1 confocal spinning-disk module (Nikon, Tokyo, Japan) and a 40X oil-immersion objective. The acquisition was done using Metamorph (Molecular Devices, San Jose, CA), at a 6 to 10 minutes acquisition rate. The focus was done on the basal plane of the cells and the microscope's hardware autofocus was used to ensure the absence of defocusing.

All the other experiments were done with a wide-field inverted microscope (Olympus, Tokyo, Japan) using a 10X air objective. Phase contrast and GFP fluorescence images were acquired using Metamorph (Molecular Devices, San Jose, CA), at a 6 to 12 minutes acquisition rate. An image of the labelled patterns was done at least at the beginning of the experiment to allow further alignment.

\section*{Polarity measurements} 

\begin{figure}[ht!]
\centering
\includegraphics[width=\textwidth]{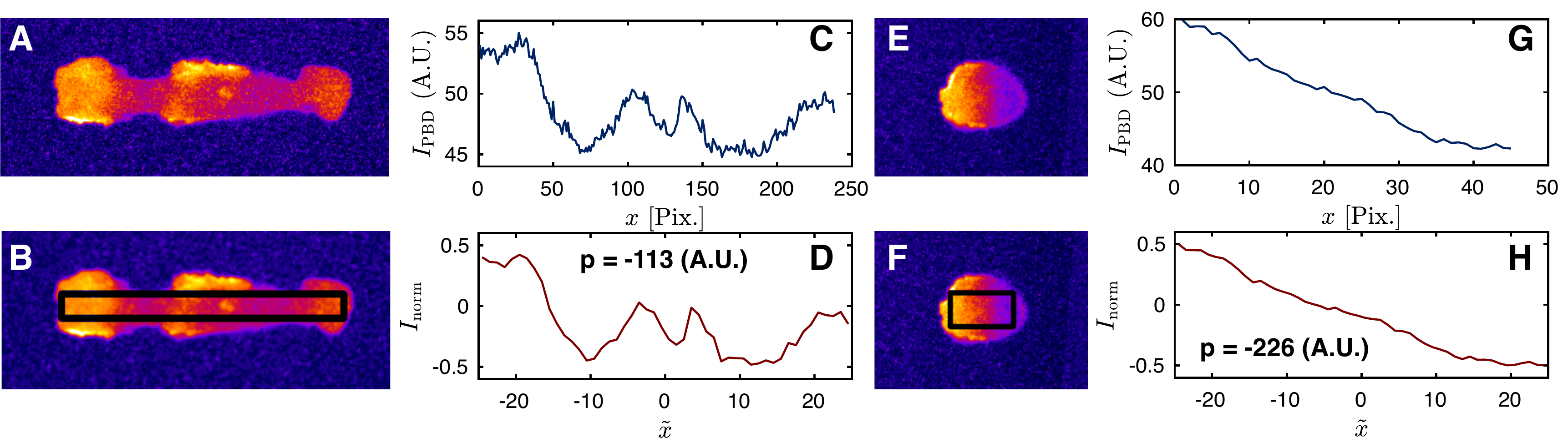}
\caption{{\bf Polarity measurement} --- (A,E) Raw snapshots of the PBD-YFP signal. (B,F) Overlaid rectangle ROI used to measure the polarity. (C,G) PBD-YFP signal along the ROI length, averaged in the transverse direction. (D,H) Same as (C,G) interpolated on normalized centered intervals both for the $x$-coordinate and the signal itself, with the corresponding computed values of $p$ in arbitrary but consistent units (here, $p<0$ in both examples, denoting a cell that is oriented to the left).}
\label{fig:supp_polarity}
\end{figure}

For polarity measurements, the images were first visually inspected using Fiji to locate cell doublets, then the images were rotated, cropped and stitched using in-house programs in Fiji and Matlab (The Mathworks, Natick, MA). The polarities were retrieved both using a semi-automated and a manual procedure due to the intrinsic heterogeneity in PBD signal. The semi-automated measurements involved drawing a rectangular ROI (Fig.\,\ref{fig:supp_polarity}B,F) at each frame for each cell in order to deal with (i) spurious, high intensity signals found mainly at the cell edges (Fig.\,\ref{fig:supp_polarity}A) and (ii) the difficulty of precisely detecting cell edges. The PBD-YFP intensity was then averaged in the direction transverse to the line (Fig.\,\ref{fig:supp_polarity}C,G) and normalized so as to range between -0.5 and 0.5. Finally, the $x$-coordinate along the cell was interpolated on a normalized symmetric coordinate $\tilde{x}$ (Fig.\,\ref{fig:supp_polarity}D,H) and the polarity was computed as the first moment of the normalized PBD signal along those symmetric coordinates, yielding the values shown in Fig.\,1D. Of note, the kymograph of cell position shown in Fig.\,1C does not exactly represent the cell edges, but rather the boundaries of the rectangular ROI, which is contained within the cell edges.

The PBD signal heterogeneity greatly limited the efficacy of the semi-automated protocol described above and led to the correct analysis of a poor number of cells. Thus, for the quantitative measurements of configurations statistics (shown in Fig.\,1E), we manually assessed the polarity of single cells, based on the visual inspection of both PBD-YFP and transmission images. The combination of the information contained in those images (PBD peaks on the edges, cell shape, lamellipodium visible in transmission) allowed us to non-ambiguously determine the -- binary -- cell polarities in most of the frames, while ambiguous data points were simply discarded.

\section*{Estimation of the relative strength of the alignment interactions} 

The evolution of the spins is governed by an equilibrium process. From the measure of the probabilities for each spin doublet configurations (see Fig.\,1E), we can estimate the value of the relative strength of the alignment interactions as follows. We denote the four possible configurations of spin doublets: $A (\rightarrow \rightarrow)$, $B (\leftarrow \leftarrow)$, $C (\leftarrow \rightarrow)$ and $D (\rightarrow \leftarrow)$. The respective probabilities of these configurations are given by 
\begin{equation}
p_A = \frac{e^{\beta}}{Z},~ p_B = \frac{e^{\beta}}{Z}, ~ p_C = \frac{e^{2\alpha - \beta}}{Z}, ~ p_D = \frac{e^{-2\alpha - \beta}}{Z}
\end{equation}
where $\alpha$ and $\beta$ are expressed in units of the spin temperature and $Z$ is the partition function. As we do not have access to the partition function, we take ratios of these probabilities to obtain the following final expressions 
\begin{align}
\alpha &= \frac{1}{4} \log (p_C/p_D) \\
\beta &= \alpha + \frac{1}{2} \log(p_A/p_C)
\end{align}
We obtain $\alpha \approx 0.59$ and $\beta \approx 0.31$, i.e. $\alpha / \beta \approx 2$. 

\section*{Image analysis}

\subsection*{Analysis of isolated trains} The line patterns were first detected using an in-house macro allowing either automatic thresholding or semi-automatic drawing. After rotation and stitching of the images using in-house macros, the phase contrast images were binarized using a simple thresholding approach with adapted smoothing, dilatation-erosion steps and filters on both the size and circularity of the detected objects. This way, the pixels occupied by cohesive trains could be detected. The GFP pictures were then treated using Imaris (Belfast, UK) to get the trajectories of single nuclei, with a manual correction step allowing to get experiment-long clean trajectories. Further analysis used a combination of those two datasets: tracked position of the single nuclei and position and extension of cohesive trains.
\\
\subsection*{Analysis of ring geometries} The analysis of ring experiments followed the same spirit but with slightly different procedures. The detection of the rings was done using in-house macros performing template matching; the images were then cropped to get single rings of all given diameters in different movies. The train detection was done as previously, but the nuclei detection relied on an in-house macro based on the \verb!Find Maxima! function of ImageJ\,\cite{dalessandro-natphys-2017}, and subsequent tracking was done using the \verb!track.m! function in Matlab (\url{http://site.physics.georgetown.edu/matlab/}). This procedure induced -- low rate -- random errors in the trajectories but allowed much larger outputs in terms of number of cells analyzed. All quantities, in particular the velocities, were projected on the orthoradial direction to get a one-dimensional dataset. In parallel, we performed particle-image velocimetry (PIV) using MatPIV (\url{https://www.mn.uio.no/math/english/people/aca/jks/matpiv/}) on the phase-contrast images. The velocity field, we obtained, was similarly projected onto the orthoradial direction, and the signal was filtered using the trains location data to remove spurious velocities outside of the areas covered by cells (see Fig.\,3E of the main text). We thus obtained the cell velocities from two independent protocols: a discrete Langrangian method and a continuous Eulerian method.
\\
\subsection*{Velocity autocorrelation functions} In both isolated trains and rings, the velocity autocorrelation functions were computed in direct space for each cell $i$ as follows
\begin{equation}
C_i(\Delta t) = \langle \mathbf{v}_i(t) \cdot \mathbf{v}_i(t+\Delta t)\rangle_t.
\end{equation}

The autocorrelation function $C_i$ was then normalized to obtain $c_i(\Delta t) = C_i(\Delta t)/C_i(0)$ and for each cell number $N$, the average $c_\mathrm{avg}(\Delta t)$ was taken over all cells belonging to a train of (initially) $N$ cells. The persistence time $\tau_v$ was defined as the first $\Delta t$ this average function decayed below $e^{-1}$. The error in $\tau_v$ was obtained by applying the same threshold detection to the functions $c_{\mathrm{avg}}\pm c_{\mathrm{sem}}$ where $c_{\mathrm{avg}}$ and $c_{\mathrm{sem}}$ are respectively the average and standard error of the mean of $c_i$ for a given $N$.
\\
\subsection*{Polar order parameter} The polar order parameter was computed on single rings with two different definitions based on the two available velocity data sets. When using individual cell tracking data, we computed the polar order parameter as follows
\begin{equation}
\mathbf{p}_\mathrm{tracking}(t) = \frac{\sum\limits_{i=1}^N \mathbf{v}_i(t)}{\sum\limits_{i=1}^N |\mathbf{v}_i(t)|},
\end{equation}
where $\mathbf{v}_i(t)$ denotes the velocity of cell $i$ at time $t$ in a ring of $N$ cells. When using the velocity field obtained via PIV, we defined the polar order parameter as 
\begin{equation}
\mathbf{p}_\mathrm{PIV}(t) = \frac{\sum\nolimits_{i\in\mathbb{T}(t)} \tilde{\mathbf{v}}_i(t)}{\sum\nolimits_{i\in\mathbb{T}(t)} |\tilde{\mathbf{v}}_i(t)|},
\end{equation}
where $\tilde{\mathbf{v}}_i(t)$ denotes the velocity measured on coarse-grained ``pixel'' $i$ at time $t$ and $\mathbb{T}(t)$ is the ensemble of pixels covered by cells at time $t$. We checked that those two definitions were consistent. In what follows, we mostly used $\langle|\mathbf{p}|\rangle_\mathrm{PIV}$ -- while checking again that the subsequent results were not affected by the definition of $\langle |\mathbf{p}|\rangle$, where $\langle \cdots \rangle$ denotes an average over configurations. To plot the steady-state values of $\langle |\mathbf{p}|\rangle$ against $N$, we waited for $t\simeq30\,\hour$ for the system to stabilize and reach stationarity. We then averaged the $\langle |\mathbf{p}|\rangle$ value for all $t\geqslant30\,\hour$ in individual rings, then binned the data obtained over number of cells $N$ (using 6 bins). We then computed both average and standard deviation over these bins, as is shown in Fig.\,3G of the main text for rings of diameter $D=400\,\micro\meter$.
\\

\section*{Polar order parameter for random configurations}

To compute the value of the polar order parameter expected for random configurations (see Fig.\,3G in the main text), we consider a set of $N$ cells of which $n_+$ have a positive polarizations and $n_-$ have a negative polarization. The global polarization of the system is given by 
\begin{equation}
p(n_+, n_-) = \frac{n_+ - n_-}{N}
\end{equation}
As $N = n_+ + n_-$, we can characterize the polarization using solely the number of negative polarities, which we denote $n$ and write
\begin{equation}
p_n = \frac{N - 2n}{N}
\end{equation}
If we consider that each cell polarization is assigned randomly, we can write that 
\begin{equation}
\langle |p_N| \rangle = \sum_{n=0}^N a_n \left| p_n \right|
\end{equation}
where $a_n$ is the probability to randomly pick $n$ cells (those with negative polarities) out of the $N$ cells given by the binomial coefficient
\begin{equation}
a_n = \frac{1}{2^N} {{N}\choose{n}}
\end{equation}
This leads to 
\begin{equation}
\langle |p_N| \rangle = \frac{2}{2^N N} {{N}\choose{1+\lfloor N/2 \rfloor}} \left[ 1 + \left\lfloor \frac{N}{2} \right\rfloor \right]
\end{equation}

\begin{figure}[t!]
\centering
\includegraphics[width=\textwidth]{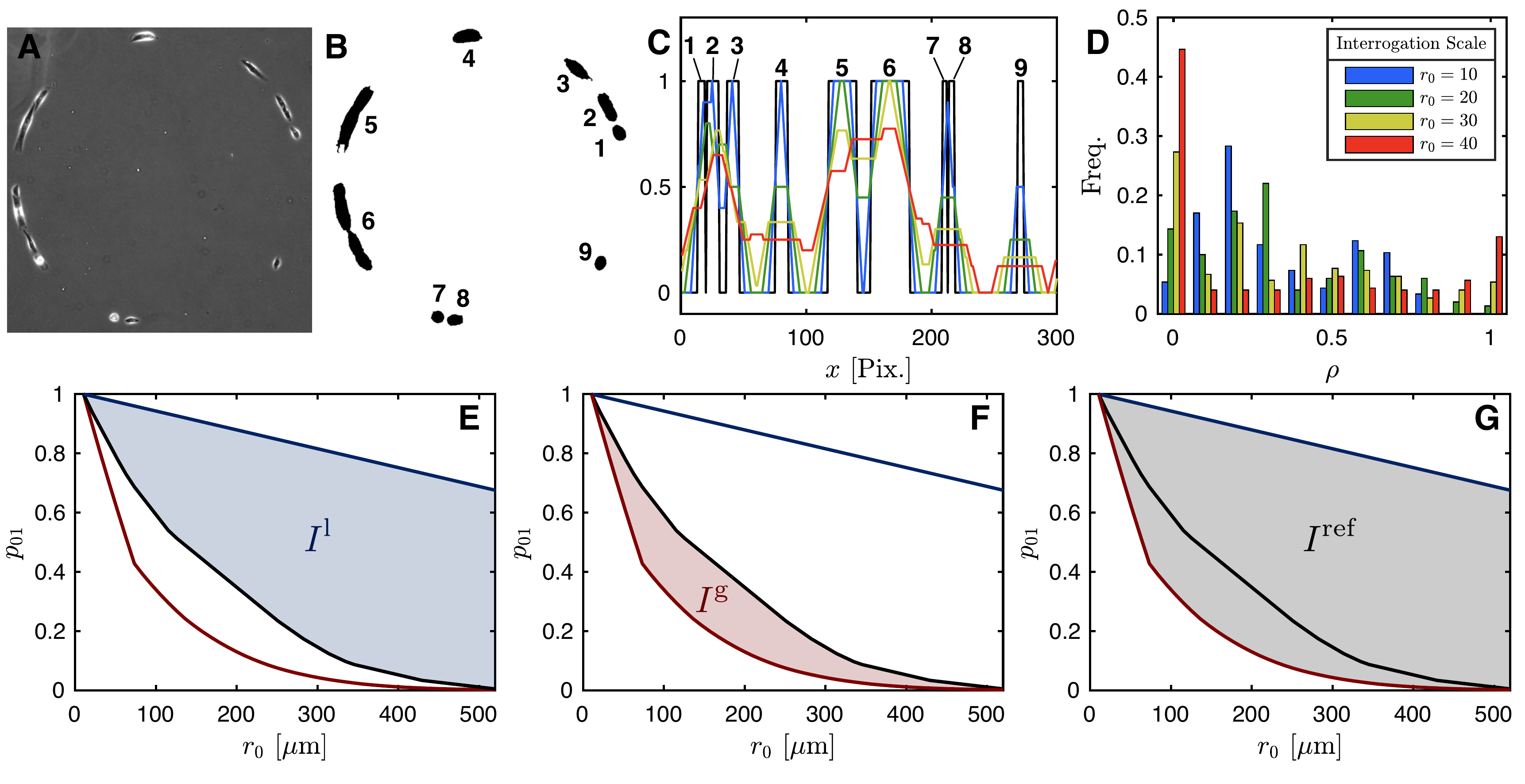}
\caption{{\bf Computation of the clustering index} ---  (A) Snapshot of a ring of diameter $D=1000\,\micro\meter$ with partially clustered cells. (B) Binarized image of the cell clusters; we index from 1 to 9 the cell clusters (``trains'') we observe. We refer to this index on the cell profile in panel (C). (C) Cell profile: raw (black) and smoothed at various interrogation scales (same colors as in panel (D)). The indices refer to the trains seen in panel (B). (D) Histogram of the ``density'' values obtained by smoothing the cell profile at various scales as in panel (C). (E-G) $p_{01}$, the frequency of 0s and 1s in the histogram obtained in (D), as a function of the interrogation scale $r_0$. The experimental curve obtained from the image in (A) is shown here in black, the corresponding simulated ideal cases (same cell number and ring diameter) are shown in blue (liquid) and red (gas). The integrals used to define the distance to both ideal cases are shown by the shaded areas in (E) (liquid), (F) (gas) and (G) (reference).}
\label{fig:supp_clustering}
\end{figure}

\section*{Clustering index} To measure the clustering index, we looked for metrics that would measure the distance of one real cell arrangement along a ring to two ideal cases: (i) a pure Poisson distribution of non-overlapping particles, and (ii) a single cluster. To that end we used the 1D binary profiles obtained from the ``train'' detection on rings (cf. above and Fig.\,\ref{fig:supp_clustering}A-B). Those profiles are simply solid clusters of 1s denoting the presence of cells and 0s denoting the absence of cells. We reasoned that a means to measure a local ``cell density'' from those binary profiles is to convolve those profiles with a typical interrogation profile (Fig.\,\ref{fig:supp_clustering}C). We choose a step function of width $r_0$ as the simplest interrogation function, which introduces the notion of interrogation scale $r_0$. Rather than relying on an optimized but arbitrary single $r_0$, we decided to study the effect of $r_0$. Clearly, the main effect of convolution is to smooth the profiles at the edges of the clusters. From a binary profile, we obtain through this procedure a continuous profile with intensity values ranging from 0 to 1. For a given cell number and a given interrogation scale $r_0$, the relative weight between intermediate values and  edge values (0s and 1s) of the intensity is highly dependent on the configuration. Indeed, a single cluster will yield the smallest number of intermediate values because it only has 2 edges, while a fully dispersed population has $2N$ edges and thus produces a lot of intermediate values (Fig.\,\ref{fig:supp_clustering}D). For a given profile, the weight of 0s and 1s along the profile smoothed at an interrogation scale $r_0$, $p_{01}$, provides a good metric to estimate the level of clustering of the configuration (Fig.\,\ref{fig:supp_clustering}E). Because this functional also depends on the covering fraction of cells, i.e. the cell number $N$, the average cell length $l$ and the ring length $L=\pi D$, we generated profiles corresponding to the two ideal configurations, for all experimental $N$ and $D$, assuming a constant $l$ based on our measurements (Fig.\,\ref{fig:supp_clustering}E). Finally, for a given real configuration, we defined its distance to the liquid-like and gas-like limit cases as the distance of its $p_{01}$ curve to those of both corresponding ideal configurations:
\begin{equation*}
d_\phi = \frac{I^\phi}{I^\mathrm{ref}}
\end{equation*}
where we define 
\begin{equation*}
I^\phi = \int_0^{r_\mathrm{max}}\!\left(p_{01}(r) - p_{01}^{\phi}(r)\right)\mathrm{d}r \quad \mathrm{and} \quad
I^\mathrm{ref}= \int_0^{r_\mathrm{max}}\!\left(p_{01}^l(r) - p_{01}^g(r)\right)\mathrm{d}r
\end{equation*}

where $r_\mathrm{max}$ is the maximal scale interrogated, $\phi\in\{\mathrm{g},\mathrm{l}\}$ denotes the ideal configuration considered, and $N$, $l$ and $L$ are implied (Fig.\,\ref{fig:supp_clustering}E-G).
Eventually, the clustering index is defined as $d_g - d_l$ and typically ranges from -0.5 to +0.5.

\begin{figure*}[b!]
\centering
\includegraphics[width=\textwidth]{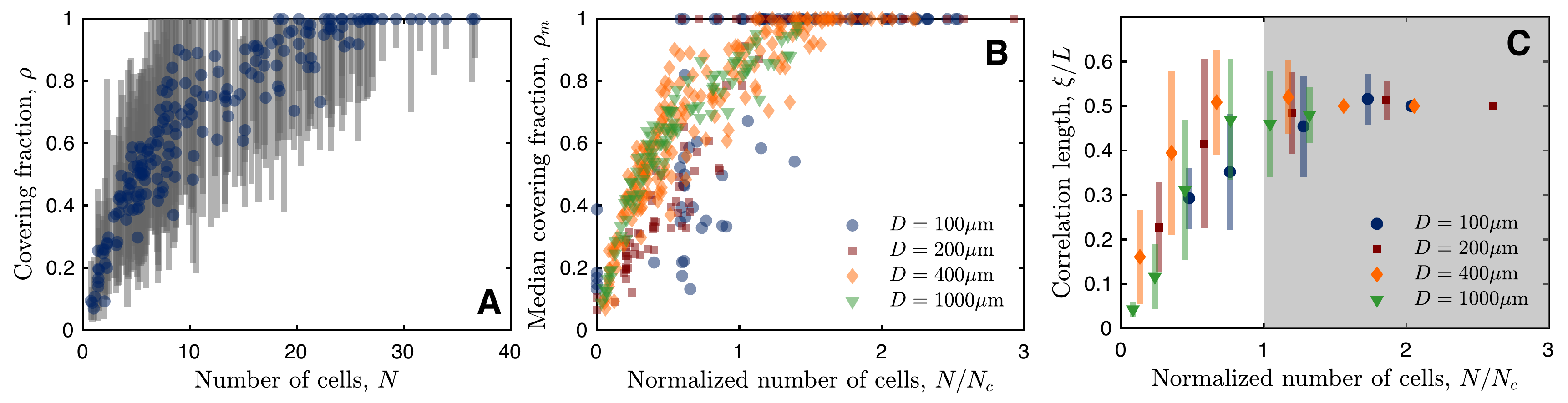}
\caption{{\bf Confluence in ring geometries} --- (A) Covering fraction $\rho$ as a function of the number of cells $N$ for a ring geometry with diameter $D=400\,\micro$m; for each experiment, blue circles represent the median covering fraction and the grey lines the minimum and maximum covering fraction observed during the experiment. We see here that $\sim20$ cells are necessary to reach confluence.  (B) Median covering fraction $\rho_m$ as a function of the number of cells normalized by the number of cells needed for confluence $N_c$ for ring geometries with diameters, $D = 100\,\micro$m (circles), $200\,\micro$m (squares), $400\,\micro$m (diamonds) and $1000\,\micro$m (triangles). (C) Velocity correlation length normalized by the ring perimeter $\xi / L$ as a function of the number of cells normalized by the number of cells needed for confluence $N_c$ for ring geometries with diameters, $D = 100\,\micro$m (circles), $200\,\micro$m (squares), $400\,\micro$m (diamonds) and $1000\,\micro$m (triangles).}
\label{fig:supp_figure1}
\end{figure*}

\section*{Critical number of cells to reach confluence}

In ring geometries, we vary the density of cells by varying the number of cells $N$ in a ring of a given diameter $D$. We measured the covering fraction $\rho$ (i.e. the fraction of the ring covered by cells) over time. As cells are treated with mitomycin C to prevent proliferation, the number of cells is conserved over time. Nevertheless, the covering fraction does vary in time. In particular, it is minimal when the cells are just plated on the fibronectin-coated ring. Over time, the cells spread on the ring and the covering fraction increases. In steady-state, we still observe significant fluctuations in covering fractions, which may be due, for instance, to the extension of lamellipodia. We show in Fig.\,\ref{fig:supp_figure1}A the covering fraction $\rho$ as a function of the number of cells $N$ for a ring geometry with diameter $D=400\,\micro$m. For each experiment, we represent the median covering fraction (blue symbols) and the range of covering fractions observed (grey lines). To measure the number of cells needed to reach confluence $N_c$, we first average the median covering fraction over bins in the number of cells $N$ and define $N_c$ as the number of cells for which this average covering fraction reaches $\rho_c = 0.85$. We proceed as follows for all ring geometries, finding the following critical number of cells to reach confluence: $N_c \approx 2$ (for $D=100\,\micro$m), $5$ (for $D=200\,\micro$m), $16$ (for $D=400\,\micro$m) and $65$ (for $D=1000\,\micro$m). In Fig.\,\ref{fig:supp_figure1}B, we show a collapse of the covering fraction $\rho$ as a function of the normalized number of cells $N/N_c$ across all ring geometries. We confirm that the validity of this parameter by showing a collapse of the velocity correlation length normalized by the ring perimeter $\xi / L$ as a function of the number of cells normalized by the number of cells needed for confluence $N_c$ across all ring geometries (see Fig.\,\ref{fig:supp_figure1}C).

\section*{Non-dimensionalization of the microscopic model}
In 1D, we nondimensionalize the Langevin equations describing our microscopic model (see Eqs.\,4 and 5 in the main text) using  $\sigma$ (the particle size) and $\varepsilon$ (the Lennard-Jones energy scale) as basic units of length and energy and as unit of time $\tau = \sigma^2 / D$, where $D = T / \zeta$ is the self-diffusion coefficient of the ABP. To do so, we introduce the self-propulsion velocity $v_0 = D F_p / T = F_p / \zeta$. We can thus define the non-dimensional P\'{e}clet number ${\rm Pe} = v_0 \tau /\sigma = v_0 \sigma / D$ which measures the ratio between the strength of the self-propulsion and thermal fluctuations.
\begin{equation}
\dot{r}_i  = \gamma F_i + {\rm Pe}\,p_i + \sqrt{2} \eta_i
\label{seq:EOM}
\end{equation}
where we defined $\gamma = \varepsilon/T$ as the ratio of the strength of the Lennard-Jones potential to the thermal fluctuations and $F_i$ as the total non-dimensionalized truncated Lennard-Jones force on particle $i$. 

We consider the polarities to be in contact with a heat reservoir temperature $T_p$; the dynamics of the spins is thus governed by flips between the values $p = +1$ and $p=-1$ with a given rate per unit of time $\mu$\,\cite{glauber-jmp-1963}. Here, we use the single spin-flip kinetic Ising model (also called Glauber dynamics). This model is defined in terms of a Markovian master equation for the probability distribution $P(p_1,\cdots,p_N,t)$, where $p_i \in \{-1,+1\}$, 
\begin{equation}
\frac{d}{dt} P(p_1,\cdots,p_N,t) = - \sum_{i} w(p_i \to -p_i) P(p_1,\cdots,p_i,\cdots,p_N,t) + \sum_{i} w(-p_i \to p_i) P(p_1,\cdots,-p_i,\cdots,p_N,t)
\end{equation}
where the transition rates $w$ are proportional to $\mu$. The transitions rates satisfy the detailed balance condition 
\begin{equation}
\frac{P_0(p_1,\cdots,-p_i,\cdots,p_N)}{P_0(p_1,\cdots,p_i,\cdots,p_N)} = \frac{w(p_i \to -p_i)}{w(-p_i \to p_i)}
\end{equation}
where the equilibrium distribution $P_{0}(p_1,\cdots,p_N) = (1/Z) \exp(-U_p/T_p)$ with $T_p$ the spin temperature, $Z$ is the partition function and $U_p$ the polarity-polarity interactions which are governed by the following Hamiltonian 
\begin{equation}\label{seq:ndUp}
U_p = \begin{cases} 
- \beta p_i \cdot p_j   - \alpha (p_i - p_j) \cdot n_{ij} & r_{ij} \le r_c \\
0 & r_{ij} > r_c
\end{cases}
\end{equation}
where $n_{ij} = r_{ij} / |r_{ij}|$. We can non-dimensionalize this Hamiltonian by expressing the symmetric alignment $\beta$ and asymmetric alignment $\alpha$ interaction strengths in units of the spin temperature $T_p$ (where in general, $T_p \ne T$). To non-dimensionalize the equations governing the spin dynamics in a way that is consistent with the basic units we detailed above, we express the spin flipping rate $\mu$ in units of the diffusion timescale $\tau$. The parameter $\mu$ is thus a non-dimensional number representing the ratio of the spatial diffusion timescale to the average polarity lifetime. In the limit where $\mu \gg 1$, particles will attempt to flip their polarities a large number of times in the time they require to diffuse by a distance corresponding to their size. Conversely, the limit where $\mu \ll 1$ corresponds to the limit of very persistent polarities.

\section*{Numerical methods} In 1D, simulations employed simultaneously: (i) the stochastic Runge-Kutta method\,\citep{branka-pre-1999} to solve Eq.\,\ref{seq:EOM} and (ii) the Glauber algorithm to solve the dynamics of the polarizations\,\citep{glauber-jmp-1963}. In all simulations, we use periodic boundary conditions and initial conditions are chosen so that the particles form a single drop with randomly chosen polarizations, $p_i = \pm 1$. The maximum timestep used is $\delta t = 0.5 \times 10^{-5}$ and we run simulations for $200 \tau$.

In our implementation of spin-flip dynamics, each particle is endowed with an internal clock. The lifetime of a given polarity is drawn from an exponential distribution with parameter $\mu $. At each timestep of the simulation, we start by updating the polarities before moving the particles. To do so, we decrease the clocks of all particles by $\delta t$, list (in order) the particles whose clocks have timed out and update them as follows: we first calculate the energy difference $\Delta E$ (in units of the temperature) resulting from the flipping of the polarity of the particle and the transition probability $P_t = 1/(1 + e^{\Delta E})$; we accept the flip if $X < P_t$ where $X$ is a uniformly distributed random number between $0$ and $1$. We then draw from the exponential distribution a new lifetime for the polarity of the particle.

\section*{Hydrodynamic Theory}
In this supplementary section, we detail the hydrodynamic theory for motile particles with asymmetric aligning interaction and its connection to the microscopic model. As discussed in the main text, contact inhibition of locomotion (CIL) can be accounted for via a potential that explicitly breaks the symmetry under {\it independent} rotations of the polarization and space (that is characteristic of equilibrium $XY$ models and the spin interaction in the Vicsek model\,\cite{vicsek-prl-1995}; although in the Vicsek model, \emph{motility} breaks this symmetry) and has symmetry only under \emph{joint} rotation of both space and polarization. The simplest model Hamiltonian with this symmetry, discussed in the main text, is 
\begin{equation}
{\cal H} = \sum_{i,j} U_r(\rv_i,\rv_j) + U_p(\rv_i,\rv_j,\pv_i,\pv_j).
\label{seq:hamiltonian}
\end{equation}
with
\begin{equation}\label{seq:Ur}
U_r = \begin{cases} 
4\varepsilon \left[ \left(\sigma/r_{ij}\right)^{12} - \left(\sigma/r_{ij}\right)^{6} \right] & r_{ij} \le r_c \\
0 & r_{ij} > r_c
\end{cases}
\end{equation}
and 
\begin{equation}\label{seq:Up}
U_p = \begin{cases} 
- \beta \pv_i \cdot \pv_j   - \alpha (\pv_i - \pv_j) \cdot {\bf n}_{ij} & r_{ij} \le r_c \\
0 & r_{ij} > r_c
\end{cases}
\end{equation}
where ${\bf n}_{ij}=\rv_{ij}/r_{ij} $ with $\rv_{ij} = \rv_{i} - \rv_{j}$, $r_{ij} = |\rv_{ij}|$, $r_c$ defines the range of interaction and $\sigma$ the particle size. The term with the coefficient $\beta$ is an $XY$ or {\it Vicsek-like} alignment term which is invariant under \emph{independent} rotations of space and polarity vectors, while the latter term with the coefficient $\alpha$ explicitly breaks the $U(1)\times U(1)$ symmetry of $U_p$ to one under $U(1)$ i.e., under \emph{joint} rotations of both space and polarity vectors. This symmetry breaking is analogous to the difference between the symmetries of the $XY$ model and liquid crystals in equilibrium. While in the Vicsek model, spin motility does break the symmetry under independent space/polarity rotations, this distinction is still important\,\cite{dadhichi-jsmte-2018} -- it implies that the ``active'' motility in the Vicsek model actually breaks \emph{two} symmetries: time-reversal and independent rotations of polarity and space, which is important, for instance, for understanding time-reversal properties and entropy production in such models\,\cite{dadhichi-jsmte-2018}. At the simplest level, the symmetry under independent rotation of space and spin of the interaction in the Vicsek model allows the interaction to be interpreted as \emph{velocity} alignment of \emph{isotropic} but \emph{motile} particles as well as alignment of polarity of polar motile particles, i.e. in those cases, ${\bf p}_i$ can be interpreted both as the polarity as well as the velocity of a particle. In this case, the potential should be even under time-reversal; however, note that if only the $\beta$ term was present in $U_p$, this condition would have been satisfied for both transformations: (i) ${\bf p_i}, {\bf p}_j\to -{\bf p_i}, -{\bf p}_j$ under $t\to-t$ (leading to the interpretation of ${\bf p}_i$ as the velocity) and (ii) ${\bf p_i}, {\bf p}_j\to {\bf p_i}, {\bf p}_j$ under $t\to t$. The presence of the $\alpha$ term, which is allowed in equilibrium polar liquid crystal models as well, breaks this symmetry and conclusively identifies ${\bf p}_i$ as the \emph{polarization} of individual particles and not their velocities.

We now write the \emph{active} dynamics of the motile particles:
\begin{align}
\zeta \dot{\rv}_i & = - \frac{\partial {\cal H}}{\partial \rv_i} + F_p \pv_i + \sqrt{2T\zeta} \bm{\eta}_i  \label{seq:overdampedR}\\
\zeta_p \dot{\pv}_i & = ( \mathbb{1}- \pv_i \pv_i^{\intercal}) \cdot \left[ - \frac{\partial {\cal H} }{\partial \pv_i} + \sqrt{2T_p\zeta_p}\bm{\xi}_i \right] \label{seq:overdampedP}
\end{align}
where $\zeta$ is the friction coefficient, $\zeta_p$ is the rotational viscosity, $T,T_p$ are the translational and polarization temperatures. There are \emph{two} sources of activity in this model: (i) motility, which enters Eq.\,\ref{seq:overdampedR} as a force $\propto {\bf p}_i$ which is not derived from a potential, and (ii) the difference between $T$ and $T_p$ which, in effect, implies that the positional and angular dynamics are connected to different baths.

We now construct a continuum version of this dynamics involving coarse-grained density and polarization fields. This ``hydrodynamic'' theory, which we will construct for arbitrary dimensions, will be valid for \emph{all} models with the same symmetry class as the microscopic model introduced above, and not only this particular microscopic model. Hydrodynamic theories are generally constructed from microscopic models using a closure scheme for the angular moments of the marginalized single-particle probability distribution function $P({\bf r}_i, {\bf p}_i, t)$, which is only strictly valid when the degree of incipient ordering in the system is small\,\cite{bertin-pre-2006}. Here, we will take a more phenomenological, less rigourous but somewhat more intuitive approach. We first define the hydrodynamic fields (i.e. those fields whose relaxation rate at zero frequency vanishes) -- these are generally of three kinds: conserved variables, transverse fluctuations of broken continuous symmetry and order parameters at a critical point. Here, the only conserved quantity is the number of particles, which means the density is a conserved quantity:
\begin{equation}
\rho({\bf r},t)=\left[\sum_{i}\delta({\bf r}-{\bf r}_i)\right]_c
\end{equation}
where $[]_c$ denotes coarse-graining using a coarse-graining kernel and where we consider unit mass particles. The second hydrodynamic field is the polarization vector or the polar order parameter. In two or higher dimensions, its fluctuations transverse to the ordering direction in the polarized phase are hydrodynamic. In one dimension, which we will finally consider here, the polar vector has an Ising symmetry and only breaks a discrete symmetry. Therefore, it is \emph{not} strictly a hydrodynamic variable in the ordered phase; however, it will still be a hydrodynamic variable at the order-disorder critical point. The polarization field is defined via
\begin{equation}
\rho({\bf r},t){\bf p}({\bf r},t)=\left[\sum_{i}{\bf p}_i\delta({\bf r}-{\bf r}_i)\right]_c.
\end{equation}
Note that although ${\bf p}_i$ is a unit vector, $|{\bf p}|\neq 1$.
Finally, even though it is not a hydrodynamic variable in this adsorbed system, we will initially retain, and then eliminate, the momentum density:
\begin{equation}
\rho({\bf r},t){\bf v}({\bf r},t)=\left[\sum_{i}{\bf v}_i\delta({\bf r}-{\bf r}_i)\right]_c.
\end{equation}

The microscopic potential can be phenomenologically coarse-grained to yield an effective free-energy in terms of the variables $\rho$ and ${\bf p}$. We do not perform this procedure here in detail and instead only provide arguments about its general form. The microscopic spin potential has two parts -- the first part $-\beta{\bf p}_i\cdot{\bf p}_j$ generically yields the usual Landau free energy\,\cite{gennes-book-1993}
\begin{equation}
F_p=\int d{\bf r}\left[\frac{A}{2}p^2+\frac{B}{4}p^4+\frac{K}{2}(\nabla{\bf p})^2\right]
\end{equation}

where all terms, in particular the parameter $A$ that controls the isotropic-polar transition, can be functions of the density. In particular, $A(\rho_c-\rho)$ turns negative at a critical density $\rho_c$ signalling an instability of the isotropic phase in which $\langle{\bf p}\rangle=0$. Note that the $(\nabla {\bf p})^2$ is a shorthand notation for the classical Frank elasticity in the one constant approximation, which comprises splay and bend contributions \cite{gennes-book-1993}, i.e. $(\nabla {\bf p})^2 = (\nabla \cdot {\bf p})^2 + |\nabla \times {\bf p}|^2$. The asymmetric exchange part of the microscopic potential, $-\alpha({\bf p}_i-{\bf p}_j)\cdot({\bf r}_i-{\bf r}_j )/|{\bf r}_i-{\bf r}_j |$ \emph{explicitly} breaks the invariance under \emph{independent} rotations of space and spin (characteristic of equilibrium $XY$ models) to one under \emph{joint} rotations of space and spin that characterizes equilibrium liquid crystals. To understand its consequences for the effective free energy, we note that a discrete approximation for the scalar $\nabla\cdot{\bf p}$ can be written as $({\bf p}_i-{\bf p}_j)\cdot({{\bf r}_i-{\bf r}_j})/{|{\bf r}_i-{\bf r}_j|}$ (see\,\cite{dhakal-pre-2010}). This is multiplied by $\Theta(|{\bf r}_i-{\bf r}_j|-r_c)$ where $r_c$ is the cut-off scale for the interaction. When this is integrated over a coarse-graining volume, this yields the local density $\rho({\bf r},t)$. Therefore, we obtain a free energy contribution explicitly coupling density and polarization of the form
\begin{equation}
F_{\rho p} =-\int d{\bf r} \bar{\alpha}\delta\rho\nabla\cdot{\bf p}
\end{equation} 
where the coefficient $\bar{\alpha}$ is proportional to $\alpha$ and $\delta\rho$ is the deviation of the density from a steady-state, spatially homogeneous density $\rho_0$ (a free-energy $\propto\rho_0\nabla\cdot{\bf p}$ with a spatially homogeneous density $\rho_0$ is a pure divergence and hence yields only a boundary term which is omitted). This is the spontaneous splay term whose effect has been studied in the context of equilibrium polar liquid crystals\,\cite{kung-pre-2006}. Invariance under joint rotations of space and spin also allows for other terms in the free energy, not present in theories of magnets, such as $p^2\nabla\cdot{\bf p}$.

The positional part of the potential, $U_r$, generically yields a free energy of the form
\begin{equation}
F_\rho=T\left[\int d{\bf r} \rho\ln\rho-\frac{1}{2}\int \int d{\bf r}d{\bf r'} C({\bf r}-{\bf r}')[\rho({\bf r})-\rho({\bf r}')]^2\right]\approx\int d{\bf r}\left[ U(\rho)+\frac{\kappa}{2}(\nabla\rho)^2\right]
\end{equation}
as in standard equilibrium theories of fluids where $C({\bf r}-{\bf r}')$ is the two-point, equal-time density correlation function for the fluid. The final expression, in which $U(\rho)$ is the internal energy density and $\kappa$ is the interface energy constant, may be obtained by Taylor expanding $\rho({\bf r}')$ about $\rho({\bf r})$. Since we are interested in model independent properties, we will take a simple phenomenological form for $U(\rho)$ $\propto (A_c/2)(\delta\rho)^2+(B_c/4)(\delta\rho)^4$ which is a function of the density fluctuations $\delta \rho$ about a steady-state density $\rho_0$.

Before proceeding to describe the active dynamics of the density and the polarization fields, we briefly discuss the effect of the free energy term induced by the asymmetric interaction in equilibrium. The free-energy $F=F_p+F_{\rho p}+F_\rho$ can be rewritten as
\begin{equation}
F=\int d{\bf r}\left[\frac{A}{2}p^2+\frac{B}{4}p^4+\frac{A_c}{2}\left(1-\frac{\bar{\alpha}^2}{A_c K}\right)(\delta\rho)^2+\frac{B_c}{4}(\delta\rho)^4+\frac{K}{2}\left(\nabla\cdot{\bf p}-\frac{\bar{\alpha}}{K}\delta\rho\right)^2+\frac{K}{2}(\nabla\times{\bf p})^2+\frac{\kappa}{2}(\nabla\delta\rho)^2\right]
\end{equation}
It is then clear that for $\rho_0 > \rho_c$ when $d\geq 2$  (with $d$ denoting the dimensionality of both the order parameter and the real space), a homogeneous uniformly polarized phase is unstable when $\bar{\alpha}^2>A_c K$\,\cite{kung-pre-2006}. Note that here, unlike in spin models which are invariant under independent rotations of space and spins, the space and spin dimensionalities are equal by construction, as enforced by the $\alpha$ term in $U_p$. However, it is crucial to note that, in the \emph{disordered} phase, the polarization fluctuations are massive and can be integrated out leaving behind a free energy only in terms of $\delta\rho$. A homogeneous \emph{disordered} phase is not destabilized by the $\bar{\alpha}$ term -- indeed, the cost in terms of density fluctuations is simply $(A_c/2)\delta\rho^2$ in this case (to lowest order in gradients). When the order parameter has an \emph{Ising} symmetry (instead of a continuous symmetry) which we expect to be the case for our one-dimensional channel geometry, the (scalar) polarization fluctuations are massive in \emph{both} the ordered and the disordered phases. The polarization fluctuations in the \emph{ordered} phase can be completely integrated out in this case, with $\delta p\sim\delta\rho$ and the $\bar{\alpha}$ term does not destroy a homogeneous polar phase with an Ising symmetry. Of course, the proliferation of domain walls destroys long-range polarization at finite temperature (as in the 1D Ising model), leading to a finite correlation length of the polarization field (defined for $\bar{\alpha}=0$); this discussion is only valid below this lengthscale.

We now construct the phenomenological \emph{active} dynamics of $\rho$, ${\bf p}$ and ${\bf v}$. The density dynamics is described by a simple continuity equation
\begin{equation}
\partial_t\rho=-\nabla\cdot(\rho{\bf v})
\end{equation}
while the equation for the polarization field is 
\begin{equation}
\label{seq:poleq}
\partial_t{\bf p}+{\bf v}\cdot\nabla{\bf p}+\bm{\Omega}\cdot{\bf p}=-\frac{1}{\gamma^R}\frac{\delta F}{\delta {\bf p}}+\Lambda{\bf v}+\lambda{\bf p}\cdot{\bsf A}+\sqrt{\frac{2T_p}{\gamma^R}}\bm{\xi}^p
\end{equation}
where $\bm{\Omega}=(1/2)[\nabla{\bf v}-(\nabla{\bf v})^T]$ and ${\bsf A}= (1/2)[\nabla{\bf v}+(\nabla{\bf v})^T]$, $\Lambda$ and $\lambda$ are phenomenological model-dependent parameters whose values depend, ultimately, on the form of $U_p$ via the $\alpha$ term (this can be understood from the fact that these dynamical terms are present in even theories of passive liquid crystals on substrates and are only invariant under joint rotations of space and polarization), $\gamma^R$ is a phenomenological polarization damping coefficient related to $\zeta_p$ in the microscopic model and $\bm{\xi}^p$ is a zero-mean, unit-variance, Gaussian, spatiotemporally white noise. Finally, the overdamped equation for the velocity field is 
\begin{equation}
\label{seq:veleq}
\gamma{\bf v}=-v_0{\bf p}-\rho\nabla\frac{\delta F}{\delta\rho}+(\nabla{\bf p})\cdot\frac{\delta F}{\delta{\bf p}}+\nabla\cdot\left[{\bf p}\frac{\delta F}{\delta{\bf p}}\right]^A+\lambda\nabla\cdot\left[{\bf p}\frac{\delta F}{\delta{\bf p}}\right]^S-\Lambda\frac{\delta F}{\delta{\bf p}}+\sqrt{{2T}{\gamma}}\bm{\xi}^v
\end{equation}
where the phenomenological damping coefficient $\gamma$ is related to $\zeta$ in the microscopic model, the motility parameter $v_0$ is replated to the motile force $F_p$ and $\bm{\xi}^v$ is a zero mean, unit variance, Gaussian, spatiotemporally white noise. Here, as in the microscopic model, activity enters in two distinct places: (i) the active motility $v_0{\bf p}$ of the particles and (ii) the distinct temperatures in Eq.\,\ref{seq:poleq} and Eq.\,\ref{seq:veleq}. Retaining terms only to the lowest order in gradients, defining $D=\rho_0^2/\gamma$, $\bar{\mu}=[(1/\gamma^R)+\Lambda^2/\gamma]$, $\Lambda_1=\rho_0\Lambda/\gamma$ and $\Lambda_2=\rho_0\Lambda/\gamma$ and eliminating the velocity field, we obtain the following closed form equations in terms of the polarization and density fields 
\begin{equation}
\label{seq:dens_pap}
\partial_t\rho=-\nabla\cdot(v_0\rho{\bf p})+D\nabla^2\frac{\delta F}{\delta \rho}+\Lambda_1\nabla\cdot\frac{\delta F}{\delta {\bf p}}
\end{equation}
and
\begin{equation}
\label{seq:pol_eq_pap}
\partial_t{\bf p}=-\bar{\mu}\frac{\delta F}{\delta {\bf p}}-\Lambda_2\nabla\frac{\delta F}{\delta \rho}
\end{equation}
where $\bar{\mu}$ is the effective polarization relaxation rate which is related to the spin-flip rate in the simulations. We are here ignoring self-advective terms of the form ${\bf p}\cdot\nabla{\bf p}$ and ${\bf p}\nabla\cdot{\bf p}$ as they do not enter in the linear analysis we consider below. Onsager symmetry dictates that $\Lambda_1=\Lambda_2$ in equilibrium. However, in this minimal active model, which breaks detailed balance, there is no symmetry to enforce this equality and, in fact, $\Lambda_1$ and $\Lambda_2$ will renormalize independently (i.e. the corrections to $\Lambda_1$ and $\Lambda_2$ under a one-loop coarse graining would, in general, be different). These coefficients could also be distinct due to microscopic non-reciprocal interactions between spins\,\cite{dadhichi-pre-2020}. While no such interaction is introduced here in the microscopic model, an effective non-reciprocal interaction can emerge in the coarse-grained theory\,\cite{dadhichi-pre-2020}. Here, for the sake of generality, we take them to be distinct with the understanding that $\Lambda_1(\rho)-\Lambda_2(\rho)=0$ when $F_p=0$ in the microscopic model i.e., $\Lambda_1(\rho)-\Lambda_2(\rho)=g(v_0, \rho)$ where $g(v_0, \rho)\to 0$ when $v_0\to 0$. The presence of terms proportional to the density gradient in Eq.\,\ref{seq:pol_eq_pap}, which appear both from the functional derivatives of the density and the polarization, also requires some comment. In usual derivations of equations of the polarization field from the microscopic dynamics of active Brownian particles, the coefficient of the term $\propto\nabla\rho$ is $\propto v_0$, i.e., the term is purely active. This is however due to the particularity of the ABP model, in which the motility, in addition to breaking time-reversal symmetry, also breaks the symmetry under \emph{independent} rotations of space and spin. Here, this symmetry is explicitly broken by the microscopic model even in the absence of $F_p$. Therefore, here the polarization field is affected by $\nabla\rho$ even in the limit of no motility. In general, Eq.\,\ref{seq:dens_pap} and Eq.\,\ref{seq:pol_eq_pap} also contain other nonlinear terms that cannot be derived from a potential. However, since they do not affect the linear physics we consider here, we omit them for simplicity.

We now linearize Eq.\,\ref{seq:dens_pap} and Eq.\,\ref{seq:pol_eq_pap} about a disordered but homogeneous phase with density $\rho_0<\rho_c$ to lowest order in gradients and eliminate the massive polarization field which is slaved to the density field via
\begin{equation}
{\bf p}=-\left(\frac{\bar{\alpha}}{A}+\frac{\Lambda_2A_c}{\bar{\mu}}\right)\nabla\rho.
\end{equation}
Inserting this in the density equation, which to lowest order in gradients, is
\begin{equation}
\partial_t\rho=-v_0\rho_0\nabla\cdot{\bf p}+DA_c\nabla^2\rho+\Lambda_1\nabla\cdot(A{\bf p}+\bar{\alpha}\nabla\rho)
\end{equation}
we obtain a closed, linear equation for the density field alone
\begin{equation}
\partial_t\rho=\left[v_0\rho_0\left(\frac{\bar{\alpha}}{A}+\frac{\Lambda_2A_c}{A\bar{\mu}}\right)-\frac{\Lambda_1\Lambda_2A_c}{\bar{\mu}}+A_cD\right]\nabla^2\rho
\end{equation}
In equilibrium, i.e. for $v_0=0$ and $\Lambda_1=\Lambda_2$, the diffusivity is simply renormalized to $D-\Lambda_1^2/\bar{\mu}>0$, as expected, and the homogenous disordered state is linearly stable for $A_c>0$, as expected; the usual spinodal line of the liquid-gas transition is then simply given by $A_c=0$.

At this stage, it is useful to compare this analysis with the equation of motion for the density and polarization fields in\,\cite{Geyer:2019aa}. These latter equations can be recovered in our formalism by taking $\Lambda_1=\bar{\alpha}=0$ and $\Lambda_2(\rho_0)\not=0$. The condition $A_c(D+v_0\rho_0\Lambda_2/A\bar{\mu})<0$ is then equivalent to the motility induced phase separation (MIPS)\,\cite{cates-arcmp-2015} spinodal obtained in\,\cite{Geyer:2019aa},  which arises with our notations when $\Lambda_2 A_c<0$ i.e., if ${\bf p}$ is interpreted as a velocity, when the effective pressure is negative. 
In the main text, consistent with\,\cite{Geyer:2019aa}, we assumed that $\Lambda_1=0$, and that $A_c\Lambda_2(v_0) $ is a decreasing function of $v_0$ with $\Lambda_2(v_0=0)=0$. Here, we now examine the effect of $\bar{\alpha}$ on this spinodal with $\Lambda_1=0$. We see that $v_0\bar{\alpha}>0$ \emph{suppresses} the instability of the homogeneous phase i.e., a higher value of $\bar{\alpha}$ makes a transition to clusters \emph{less} likely. This is expected since, when the polarization is interpreted as a velocity field, the $\bar{\alpha}$ term is a pressure-like term. Furthermore, even the usual equilibrium spinodal line, when $A_c<0$, is shifted to a finite \emph{negative} value due to $\bar{\alpha}$. The simplest way to see this is to consider the special case where $\Lambda_1=\Lambda_2=0$, i.e., the \emph{only} nonequilibrium contribution to the dynamics arises from $v_0$. In this case, a positive $\bar{\alpha}$ shifts the spinodal from $A_c=0$ to $v_0\rho_0\bar{\alpha}/AD$ (this limit is somewhat artificial however as $\Lambda_2$ is generically expected to have a contribution from motility). 
Thus, unlike in equilibrium, the spinodal is shifted due to $\bar{\alpha}$ even in the disordered phase due to the presence of $v_0$.  

Finally, for completeness, we linearize Eq.\,\ref{seq:dens_pap} and Eq.\,\ref{seq:pol_eq_pap} in one dimension about an ordered and homogeneous phase with density $\rho_0>\rho_c$, for system-spanning clusters that are smaller than the Ising correlation length (i.e. for system sizes smaller than the Ising correlation length), away from the critical point, to lowest order in gradients:
\begin{equation}
\partial_t\rho=-\left(\frac{\bar{A}\rho_0}{2|A|}+p_0\right)v_0\partial_x\delta\rho+\left[v_0\rho_0\left(\frac{\bar{\alpha}}{2|A|}+\frac{\Lambda_2A_c}{2|A|\bar{\mu}}\right)-\frac{\Lambda_1\Lambda_2A_c}{\bar{\mu}}+A_cD\right]\partial_x^2\rho
\end{equation}
where $\bar{A}=\partial A/\partial\rho|_{\rho_0}$. The first term, $\propto\partial_x\delta\rho$, demonstrates that in a homogeneous polarized phase, density fluctuations are ballistically propagated by the polarization field, which further reinforces its velocity-like character, while the positivity of the term in the square brackets determines the stability of the homogeneous polar phase.

\vspace{0.5cm}

\section*{Supplementary Movies}
\begin{enumerate}
\item Movie accompanying the data from the MDCK PBD-YFP cell doublet presented in Fig.\,1C-D.

\item Movie accompanying the snapshots shown in Fig.\,2A --- movie of an isolated cell train with $N=3$ on a linear geometry in transmission with fluorescently tagged nuclei showing repolarization events and dynamic lamellipodia activity.

\item Movie accompanying the snapshots shown in Fig.\,2B --- movie of an isolated cell train with $N=8$ on a linear geometry in fluorescence (PBD) showing fracture of the cell train followed by repolarization events and dynamic lamellipodia activity.

\item Movie accompanying the snapshots shown in Fig.\,2C --- movie of ring geometries for various ring diameters and cell numbers showing the emergence of coordinated polarization and collective circular motion at confluence.
\end{enumerate}

\end{document}